\documentclass[letterpaper]{vldb}

\title{On Joining Graphs}
\numberofauthors{3}
\author{
	\alignauthor
	Giacomo Bergami\\
	\affaddr{University of Bologna}\\
	\affaddr{CSE Department}\\
	\affaddr{Bologna, Italy}\\
	\email{giacomo.bergami2@unibo.it}
	\alignauthor
	Matteo Magnani\\
	\affaddr{Uppsala University}\\
	\affaddr{Department of Information Technology}\\
	\affaddr{Uppsala, Sweden}\\
	\email{matteo.magnani@it.uu.se}
	\alignauthor
	Danilo Montesi\\
	\affaddr{University of Bologna}\\
	\affaddr{CSE Department}\\
	\affaddr{Bologna, Italy}\\
	\email{danilo.montesi@unibo.it}
}
\usepackage{url}
\usepackage{microtype}
\usepackage{balance}
\usepackage{adjustbox}
\usepackage{balance}
\usepackage[table]{xcolor}

\usepackage{xfrac}
\usepackage{graphicx}

\usepackage{mwe}

\usepackage{resizegather}

\usepackage{amsmath}
\usepackage{mathtools}

\usepackage{braket}

\usepackage{cancel}

\usepackage{algorithm}

\usepackage[noend]{algpseudocode}

\usepackage{comment}

\usepackage{varioref}

\usepackage{pifont}

\newcommand{\cmark}{\ding{51}}%
\newcommand{\xmark}{\ding{55}}%

\usepackage{amssymb}

\usepackage{listings}

\usepackage{caption}
\usepackage{subcaption}

\usepackage{booktabs}

\usepackage{multirow}

\usepackage[utf8]{inputenc}

\usepackage{balance}

\usepackage{csvsimple}

\lstdefinelanguage{sparql}{
	morekeywords={SELECT,CONSTRUCT,DESCRIBE,ASK,WHERE,FROM,NAMED,PREFIX,BASE,OPTIONAL,FILTER,GRAPH,LIMIT,OFFSET,SERVICE,UNION,EXISTS,NOT,BINDINGS,MINUS,GROUP,BY, HAVING, COUNT, DISTINCT,a},
	sensitive=true
}
\lstdefinelanguage{cypher}{
	morekeywords={MATCH,RETURN,WHERE,DISTINCT,WITH,CREATE,COUNT,AS,UNION,ALL,is,null,NOT,AND,OR},
	sensitive=true
}
\lstdefinelanguage{gremlin}{
	morekeywords={g,V,match,as,hasLabel,out,has,inE,outV,as,between,select,groupCount,TinkerGraph,io,IoCore,readGraph,traversal,values,in,count,is},
	sensitive=true
}
\lstdefinelanguage{biql}{
	morekeywords={CREATE,SELECT,FROM,WHERE,AS,count},
	sensitive=true
}
\lstdefinelanguage{nosql}{
	morekeywords={create,class,extends,edge,vertex,select,from,where,in,or,and,not,out,delete,insert,into,as},
	sensitive=false
}
\graphicspath{{./}{../generalfig2/}}
\newcommand{\phparagraph}[1]{\paragraph*{#1} \addcontentsline{toc}{paragraph}{#1}}
\usepackage{pifont}

\usepackage{amsthm}
\newtheorem{definition}{Definition}
\newtheorem{example}{Example}
\usepackage{threeparttable}

\algdef{SE}[DOWHILE]{Do}{doWhile}{\algorithmicdo}[1]{\algorithmicwhile\ #1}%

\usepackage[
type={CC},
modifier={by-nc-sa},
version={4.0},
]{doclicense}
\PassOptionsToPackage{hyphens}{url}\usepackage{hyperref}
\toappear{\doclicenseText\doclicenseIcon
	
	}

\begin{document}
	\maketitle
	\begin{abstract}
In the graph database literature the term ``join'' does not refer to an operator used to merge two graphs. In particular, a counterpart of the relational join is not present in existing graph query languages, and consequently no efficient algorithms have been developed for this operator.

This paper provides two main contributions. First, we define a binary graph join operator that acts on the vertices as a standard relational join and combines the edges according to a user-defined semantics. Then we propose the ``CoGrouped Graph Conjunctive $\theta$-Join'' algorithm running over data indexed in secondary memory. Our implementation outperforms the execution of the same operation in Cypher and SPARQL on major existing graph database management systems by at least one order of magnitude, also including indexing and loading time.
	\end{abstract}
	
	\section{Introduction}
Despite the term ``join'' appearing in the literature on GDBMSs, there is no counterpart of
the relational join operator over two distinct graphs. 
Recalling relational 
algebra's joins, those are defined as a composition of 
a selection predicate and a cartesian product between 
two tables.
In literature ``graph join'' describes an operation which is
neither binary (that is, involving two distinct graphs) nor involving graphs
as a whole (that is, graph paths obtained in traversal operations over one single
graph are considered instead). 	
In current literature ``join'' expresses a ``path join'' \cite{LiM03,Holzschuher,Gao} over 
path queries of arbitrary length, where specifically joins are performed over adjacent vertices \cite{Atre,Yuan,Fletcher09}.

	\begin{figure*}
		\centering
		\begin{minipage}[b]{.2\linewidth}
			\includegraphics[scale=0.7]{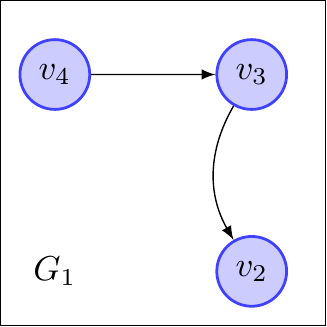}
			\subcaption{$G_1=(V_1,E_1,A_1)$}\label{fig:figmaina}
		\end{minipage}%
		\begin{minipage}[b]{.25\linewidth}
			\begin{tabular}{|c|c|c|}
				\hline
				\rowcolor{lightgray} \multicolumn{3}{|c|}{$V_1\mathtt{(User,MsgTime1)}$} \\
				\hline
				Vertex & \texttt{User} & \texttt{MsgTime1} \\
				\hline
				$v_2$ & Alice & 1\\
				$v_3$ & Bob  & 3 \\
				$v_4$ & Carl & 2 \\
				\hline
			\end{tabular}
			\subcaption{$V_1$}\label{fig:v1}
		\end{minipage}
		\quad
		\begin{minipage}[b]{.2\linewidth}
			\begin{tabular}{|c|c|}
				\hline
				\rowcolor{lightgray} \multicolumn{2}{|c|}{$E_1$} \\
				\hline
				Source & Destination \\
				\hline
				$v_3$ & $v_2$\\
				$v_4$ & $v_3$\\
				\hline
			\end{tabular}
			\subcaption{$E_1$}\label{fig:e1}
		\end{minipage}\\
		\bigskip 
		
		\begin{minipage}[b]{.2\linewidth}
			\includegraphics[scale=0.7]{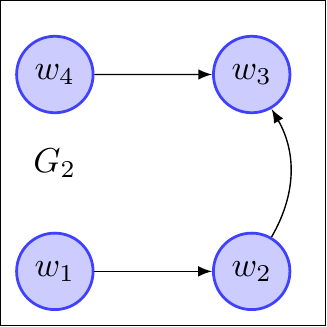}
			\subcaption{$G_2=(V_2,E_2,A_2)$}\label{fig:figmainb}
		\end{minipage}
		\begin{minipage}[b]{.25\linewidth}
			\begin{tabular}{|c|c|c|}
				\hline
				\rowcolor{lightgray} \multicolumn{3}{|c|}{$V_2\mathtt{(User,MsgTime2)}$} \\
				\hline
				Vertex & \texttt{User} & \texttt{MsgTime2} \\
				\hline
				$w_1$ & Dan & 6\\
				$w_2$ & Alice & 7\\
				$w_3$ & Bob  & 3 \\
				$w_4$ & Carl & 2 \\
				\hline
			\end{tabular}
			\subcaption{$V_2$}\label{fig:v2}
		\end{minipage}
		\quad
		\begin{minipage}[b]{.2\linewidth}
			\begin{tabular}{|c|c|}
				\hline
				\rowcolor{lightgray} \multicolumn{2}{|c|}{$E_2$} \\
				\hline
				Source & Destination \\
				\hline
				$w_1$ & $w_2$\\
				$w_2$ & $w_3$\\
				$w_4$ & $w_3$\\
				\hline
			\end{tabular}
			\subcaption{$E_2$}\label{fig:e2}
		\end{minipage}
		\caption{\textit{Dual representation of graphs as both graphs and data tables}.}\label{fig:dataexample}
	\end{figure*}
	\begin{figure}[!t]
		\centering
		\includegraphics[scale=0.7]{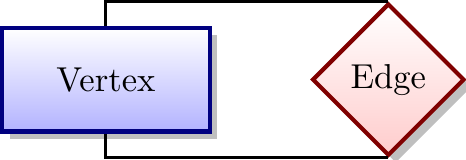}
		\caption{\textit{View of the data graph as a relational database}.}
		\label{fig:ger}
	\end{figure}

	In order to 
	reinforce such statement, we should see that the 
	so-called graph joins in graph database literature
	are always defined over a single graph \cite{SIGMOD2015Atre}
	even if there are some graph query languages, like SPARQL, that
	allow accessing multiple graph resources in a same query\footnote{See ``named graphs'' in SPARQL.}.
	In the particular case of SPARQL, the join operation is expressed as a
	join between graph paths through relational join operators \cite{SparQLExpr,SIGMOD2015Atre}
	through triple composition.
	
	The introduction of an explicit graph join operator allows to better delineate
	the problem, and hence helps finding a tailored algorithm that performs the
	join operation profitably. By doing so we obtain a specific join algorithm
	that has a running time that is lower than the time required to perform the
	same operation over current graph database query languages.
	
	In this paper we outline two contributions: 
	\begin{itemize}
		\item \textbf{Definition of the Graph Join operator}: this operation
		joins the vertices and combines the edges with different possible
		semantics. In particular
		we propose two
		different semantics, that are the \textbf{conjunctive join} and the
		\textbf{disjunctive join}. The first one forms
		an edge between two joined vertices iff such vertices were connected
		by an edge in both graphs; the second one forms an edge if the vertices
		were connected in at least one of the two graphs.

		\item \textbf{CoGrouped Graph Conjunctive $\theta$-Join}: this algorithm
		implements the graph join with the conjunctive semantics. First, we index and load into secondary 
		memory the graph data structure by associating each vertex to a specific hash value. As a last step,
		we perform a join over the operands which are stored in secondary memory by both (i)  joining only the vertices
		which have the same hash value, and (ii) linking such joined vertices according to the
		conjunctive semantics.
	\end{itemize}
	
	%
	
As a secondary outcome of the flexibility of the graph join definition,
we implement specific graph operations
even though those are not literary specified as graph joins.
\textit{Subgraph extractions from two graphs} over user
communities \cite{Berlingerio11,Boden12}
and \textit{(Unweighted) Ontology RollUp} \cite{Mabroukeh11} over a same graph are just a few examples.
	
	%
	%
	
	Section \ref{sec:gdm} outlines the details of our proposed graph data model and 
	defines the graph join operation (Section \ref{subsec:joindefs}). 
	Section \ref{sec:algos} develops both the basic version of the join definitions
	and the proposed CoGrouped Graph Conjunctive $\theta$-Join, specifically designed for
	the graph conjunctive join. Section \ref{ref:structure}
	exposes the graph data structures that are used both to store the result in primary memory 
	and the join operands in secondary memory.
	Section \ref{sec:dataset}
	describes the experiment's set-up where we generalize the subgraph extraction problem
	in order to increase the results' multiplicity.
	In Section \ref{src:conc} we draw our conclusions and outline our future works. 
	Section \ref{sec:dbqlang} provides more details on the state of the art
	of current graph database languages and on the usage of the term ``join'' on current
	graph database literature.

	
	\section{Graph Data Model}\label{sec:gdm}
	We now define a simplified Property Graph data model, which uses only the basic desired feature 
	that are required to develop a graph join operation over graphs. 
	\begin{definition}[Graph Data Model]
		A graph is defined as a triple $(V,E,A)$, where $V$ is a set of vertices, that are represented as tuples having a schema $A$. $E$ is a set of (unlabelled) edges defined as a pair of vertices in $V^2$. 
	\end{definition}
	
	Using the former definition, we can represent any graph through a
	ER diagram (Figure \ref{fig:ger}). This simplified model could 
	be even stored in a  relational database (Figure \ref{fig:dataexample}), because  each 
	vertex is an entity and the edges are the binary relations among the vertices. 
	Similar attempts have been carried out for the Property Graph model 
	\cite{preSQLGraph,SQLGraph}.
	
	\begin{example}
		Figure \ref{fig:figmaina} and \ref{fig:figmainb} represent two communication patterns between the two vertices,
		where each vertex represents a post created by an user and each edge $(u,v)$ represents that $u$ receives
		a reply from $v$. Consequently, we have:
		\begin{gather*}
		G_1=(\Set{v_2,v_3,v_4},\Set{(v_2,v_3),(v_3,v_2),(v_4,v_3)},\Set{User,MsgTime1})\qquad\qquad\quad\\
		G_2=(\Set{w_1,w_2,w_3,w_4},\Set{(w_1,w_2),(w_2,w_3),(w_3,w_2),(w_4,w_3)},\Set{User,MsgTime2})
		\end{gather*}
		Each graph 
		can be represented by two tables, one for the vertices (Figure \ref{fig:v1} and
		\ref{fig:v2}) and the other for the edges (Figure \ref{fig:e1} and \ref{fig:e2}).
	\end{example}
	
	The evidence of a viable mapping between graphs and relational databases is used in the following subsection to outline the graph join operation.
	%

	\begin{figure}[!b]
		\centering
		\includegraphics[scale=0.7]{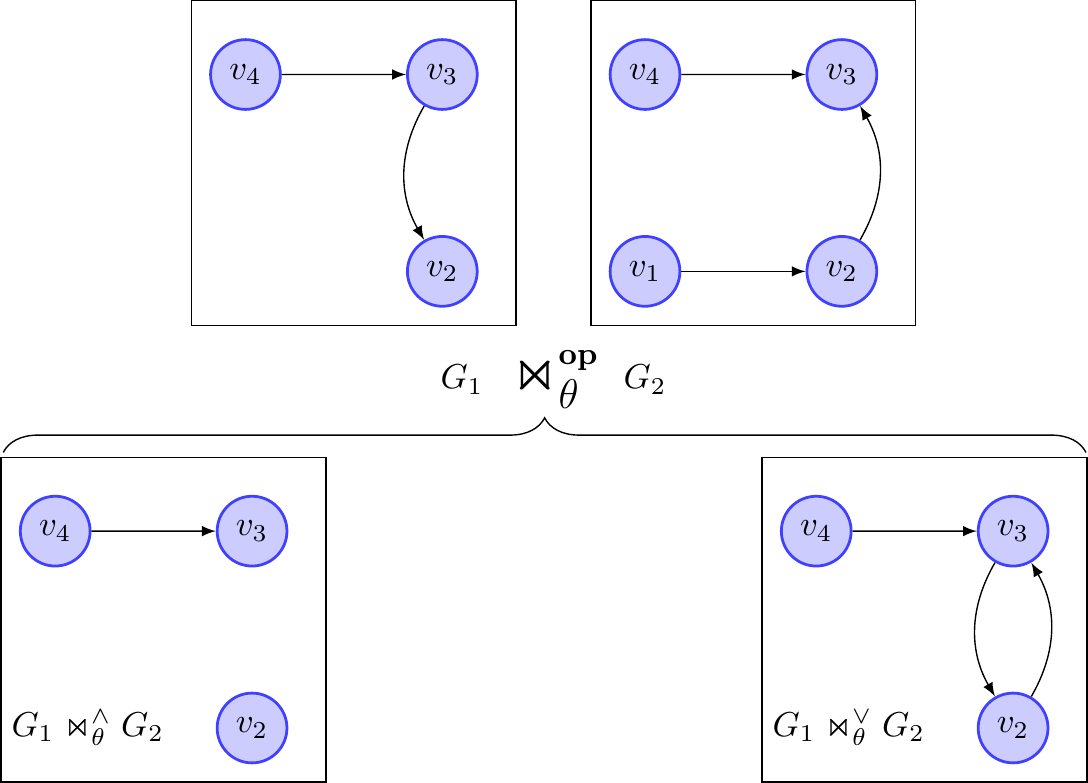}
		\caption{\textit{Given two graph with vertices with same id, hence sharing the same value, the graph conjunctive 
				join extracts the common pattern, while the disjunctive join retrieves at least one edge shared 
				among the matched vertices}.}
		\label{fig:conjdisjbasicex}
	\end{figure}
	
	\begin{figure*}[!t]
		\centering
		\begin{adjustbox}{max width=\textwidth}
			\begin{minipage}[b]{.2\textwidth}
				\includegraphics[scale=0.7]{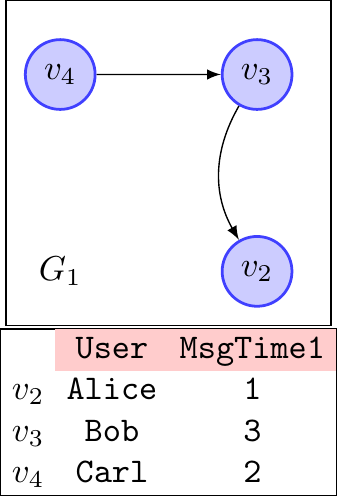}
				\subcaption{$G_1$}\label{fig:figjoing1}
			\end{minipage}
			\quad
			\begin{minipage}[b]{.18\textwidth}
				\includegraphics[scale=0.7]{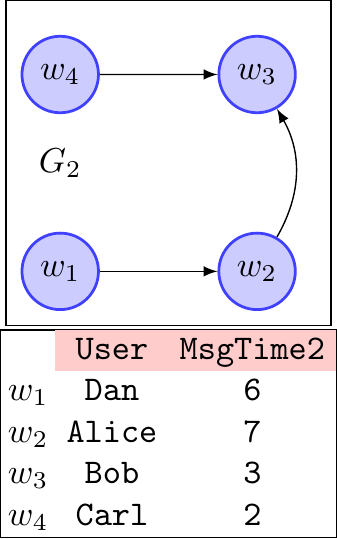}
				\subcaption{$G_2$}\label{fig:figjoing2}
			\end{minipage}
		\end{adjustbox}\\
		\bigskip

		\begin{adjustbox}{max width=\textwidth}
			\begin{minipage}[b]{.25\linewidth}
				\includegraphics[scale=0.7]{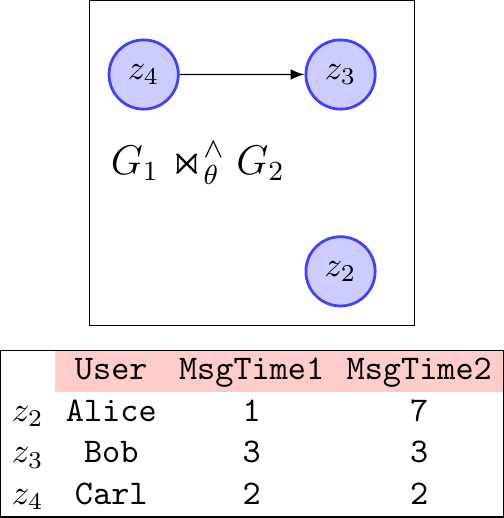}
				\subcaption{$G_1\Join_\theta^\wedge G_2$}\label{fig:figjoina}
			\end{minipage}\quad
			\begin{minipage}[b]{.25\linewidth}
				\includegraphics[scale=0.7]{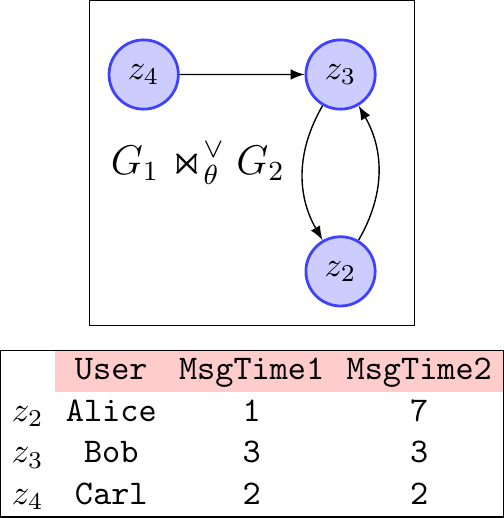}
				\subcaption{$G_1\Join_\theta^\vee G_2$}\label{fig:figjoinda}
			\end{minipage}\quad
			\begin{minipage}[b]{.25\linewidth}
				\includegraphics[scale=0.7]{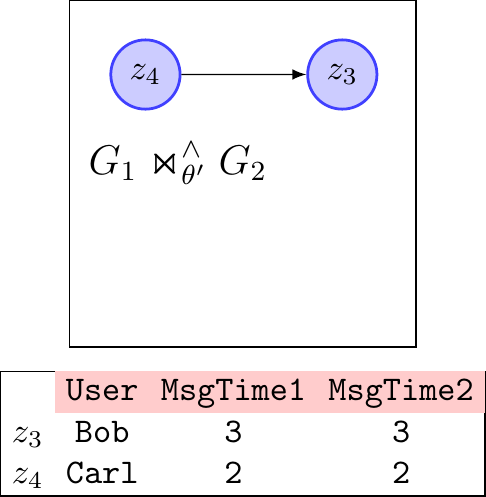}
				\subcaption{$G_1\Join_{\theta'}^\wedge G_2$}\label{fig:figjoinb}
			\end{minipage}\quad
			\begin{minipage}[b]{.25\linewidth}
				\includegraphics[scale=0.75]{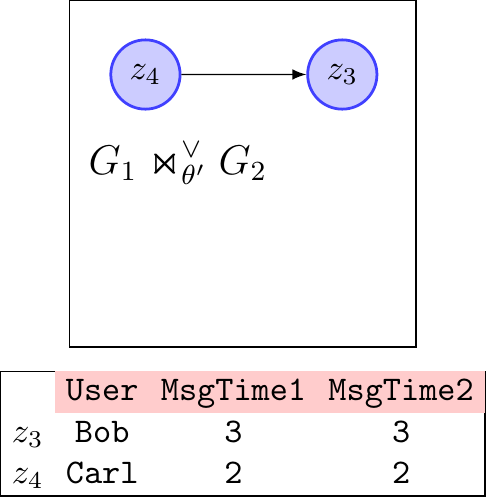}
				\subcaption{$G_1\Join^\vee_{\theta'} G_2$}\label{fig:figjoindb}
			\end{minipage}
		\end{adjustbox}
		\caption{\textit{Representation of $G_1\Join_\theta^\vee G_2$, where $\theta$ is 
				defined as ${\texttt{MsgTime1}\leq\texttt{MsgTime2}}$ and $\theta'$ is
				defined as ${\texttt{MsgTime2}\leq\texttt{MsgTime1}}$}.}
		\label{fig:dataexamplejoin1d}
	\end{figure*}

	\subsection{Graph Joins}\label{subsec:joindefs}
	We conceive graph joins as an extension of  the relational join operator.
	We propose a graph join between two graphs $G_1\Join_\theta^{\textup{\textbf{op}}}G_2$,
	where the vertices are considered as to-be-joined
	relational tuples ($V_1\Join_\theta V_2$), 
	and the resulting edges are given by combining the graph operand's edges $E_1$ and $E_2$ with a specific
	\textbf{op} semantics. The graph join operator has two parameters: 
	the $\theta$ binary predicate over the vertices and the \textbf{op}
	semantics that combines the edges from both graphs.
	This  modularity is similar to
	the graph products defined in graph theory literature 
	\cite{Hammack,ProductGraphs}, where instead of a join between vertices we have a 
	cross product. 
	\begin{definition}[General Graph $\theta$-Join]
		Given two data graphs $G_1=(V_1,E_1,A_1)$ and $G_2=(V_2,E_2,A_2)$, a \textbf{general graph 
			$\theta$-join} is defined as follows:
		\begin{gather*}
		G_1\Join_\theta^{\textbf{\textup{op}}} G_2=(V_1\Join_\theta V_2,E_{\textup{\textbf{op}}},A_1\cup A_2)
		\end{gather*}
		where $\theta$ is a binary predicate over the vertices and $\Join_\theta$ the $\theta$-join among the vertices (as tables), and 
		$E_{\textup{\textbf{op}}}$ is a subset of all the possible edges linking the vertices in $V_1\Join_\theta V_2$ expressed 
		with the \textup{\textbf{op}} semantics.
		$\Join_\theta$ is defined as the relational join among
		relations as follows:
		\begin{gather*}
		V_1\Join_\theta V_2=\Set{v\oplus v''|v\in V_1, v''\in V_2,\theta(v,v''), (v\oplus v'')[A_1]=v, (v\oplus v'')[A_2]=v''}
		\end{gather*}
		Moreover, $\oplus$ is the operation of merging two vertices (and hence, two tuples).
	\end{definition}
	If we impose that $E_{\textup{\textbf{op}}}$ contains an edge iff an edge between the
	merged vertices appears in both original graphs (and hence in both $E_1$ and $E_2$),
	we obtain a \textbf{Conjunctive Join}, that in graph theory is known as \textit{Kronecker graph product} \cite{Weichsel,Hammack}.
	In this case  $E_{\textup{\textbf{op}}}$ is defined with the ``$\wedge$'' semantics as follows:
	\begin{gather*}
	E_{\wedge}=\{(v\oplus v'',v'\oplus v''')\in (V_1\Join_\theta V_2)^2\mid (v,v')\in E_1\wedge (v'',v''')\in E_2 \}
	\end{gather*}
	Such edge semantics extracts all the shared edge patterns between the two graphs among equivalent vertices,
	where $\theta$ determines whether the two vertices are equivalent or not. Given this definition, we could provide an
	implementation for the Unweighted Ontology RollUp \cite{Mabroukeh11} operator:
	
	\begin{example}
	Given a graph $G=(V,E)$ where each edge represents a 
		``is-a'' relation,
		then $G\Join_{(v,v')\in E}^\wedge G$ produces the rollup for each ontology object $o\in V$ over which 
		``is-a'' relations are formulated. 
		E.g., if \textup{camera} ``is-a'' \textup{device} and \textup{device} ``is-a'' \textup{hardware}, then in the resulting
		graph \textup{camera-device} ``is-a'' \textup{device-hardware}
	\end{example}
	
	This
	graph join semantics is implemented in our proposed algorithm
	in Section \ref{sec:cogroup}.
	As an alternative, we would like to grasp from both graphs all the edges that are shared among the equivalent
	edges (see Figure \ref{fig:conjdisjbasicex}): this is possible with the  \textbf{Disjunctive Join}, where $E_{\textup{\textbf{op}}}$
	is defined with the ``$\vee$'' semantics as follows:
	\begin{gather*}
	E_{\vee}=\{(v\oplus v'',v'\oplus v''')\in V_1\Join_\theta V_2\mid (v, v')\in E_1\vee (v'', v''')\in E_2\}
	\end{gather*}

	In order to differentiate the two proposed graph joins, that is when the conjunctive edge semantics is used
	instead of the disjunctive one, we use $G_1\Join^\wedge G_2$ for the graph conjunctive join, and $G_1\Join^\vee G_2$
	for the graph disjunctive one.

	\begin{example}\label{ex:4}
		Let us now consider the two graphs in Figure \ref{fig:dataexample} and try to define
		the join $G_1\Join_{\texttt{MsgTime1}\leq\texttt{MsgTime2}} G_2$, where each vertex 
		value is provided inside a table. We want
		to select the communication patterns that are shared at increasing times: the resulting 
		graph is given in Figure \ref{fig:figjoina}, where each $z_i$ is defined as $v_i\oplus w_i$. Observe that
		the graph conjunctive join operation is symmetric as the one defined in relational algebra, and hence:
		\[G_1\Join_{\texttt{MsgTime1}\leq\texttt{MsgTime2}}^\wedge G_2=G_2\Join_{\texttt{MsgTime1}\leq\texttt{MsgTime2}}^\wedge G_1\]
		
		On the other hand, the result of performing the symmetric operation, that is $G_1\Join_{\texttt{MsgTime2}\leq\texttt{MsgTime1}} G_2$,
		is provided in Figure \ref{fig:figjoinb}. Observe that the graph disjunctive join operation is
		symmetric as the one defined in relational algebra, and hence:
		\[G_1\Join_{\texttt{MsgTime2}\leq\texttt{MsgTime1}}^\vee G_2=G_2\Join_{\texttt{MsgTime2}\leq\texttt{MsgTime1}}^\vee G_1\]
		
		Let us now consider the same examples provided previously but performed over the Disjunctive Join: 
		as we can see from Figure \ref{fig:figjoinda}, such implementation allows obtaining the
		edges from the two graphs and behaves as a missing data operator for the edges. As we can see
		from Figure \ref{fig:figjoindb}, whether both graphs share the same edge, the final edge is 
		not duplicated.		
	\end{example}

	\section{Algorithms}\label{sec:algos}
	In this section we design both a basic approach implementing the graph join for
	both conjunctive ($\wedge$) and disjunctive ($\vee$) semantics (Section \ref{sub:baisc}), 
	and then we outline the proposed join algorithm for the conjunctive semantics (Section \ref{sec:cogroup}),
	focusing more specifically on the equijoin case (the discussion on the less-equal join
	is postponed to Appendix \ref{app:leq}).

		\begin{algorithm}[!b]
			\caption{Basic Join, implementing both conjunctive and disjunctive join, depending on
				\textbf{op}'s definition (that can be either \textbf{and} or \textbf{or})}\label{alg:triviial}
			\begin{algorithmic}[1]
				\Procedure{Join}{$G_1,G_2,\theta,$\textbf{op}}\Comment{$G_1\Join_\theta^{\textup{\textbf{op}}} G_2$}
				\State $V\gets V_1\Join_\theta V_2; E\gets \emptyset$
				\For{\textbf{each} $l\oplus r\in V$}
				\For{\textbf{each} $ll\oplus rr\in V$}
				\If{($(l,ll)\in E_1$ \textbf{op} $(r,rr)\in E_2$)}
				\State $E\gets E\cup \{(v,v_2)\}$
				\EndIf
				\EndFor
				\EndFor
				\State \textbf{return} $(V,E,A_1\cup A_2)$
				\EndProcedure
			\end{algorithmic}
		\end{algorithm}
	
	\subsection{Basic implementation}\label{sub:baisc}
	A basic implementation of both conjunctive and disjunctive join is provided in Algorithm \vref{alg:triviial}. 
	This simple algorithm can be immediately translated in common graph query languages such
	as Cypher and SPARQL. To the best of our knowledge, no graph language
	exists where such graph join operator is designed as a specific operator and hence, the computation of
	such kind of query is not optimized, although all the graph database indexing techniques are used. 
	
	With this implementation we first perform the $\theta$ join over the graphs' vertices (line 2),
	 then we visit the graph searching for the neighbours (line 4) and then we check if the selected vertices
	satisfy the specific semantics through the \textbf{op} semantics.
	
	
	\begin{figure*}[!th]
		\centering
		\begin{minipage}[b]{\linewidth}
			\hspace*{2cm}  
			\includegraphics[scale=0.75]{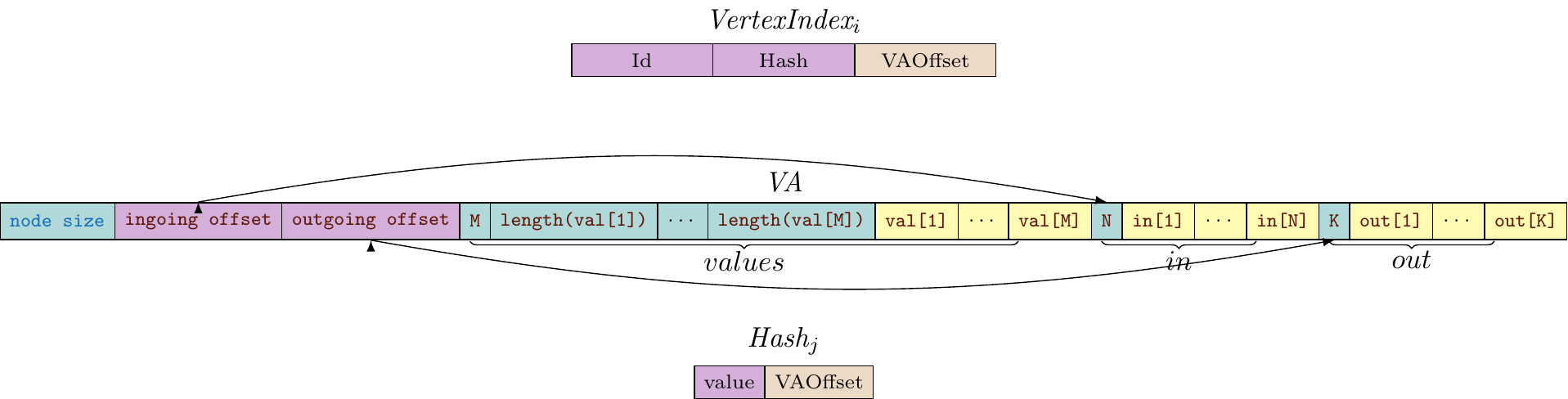}
			\subcaption{Data structures used to implement the graph in secondary memory. Each data structure represents a different file.}\label{fig:graphstructure}
		\end{minipage}\\
		\begin{adjustbox}{max width=\textwidth}
			\hspace*{-0.5cm}  
			\begin{minipage}[b]{.1\textwidth}
				\includegraphics[scale=0.68]{g1_mod_tab}
				\subcaption{$G_1$}\label{fig:figjoing1bis}
			\end{minipage} \begin{minipage}[b]{0.8\textwidth}
			\hspace{1cm}\includegraphics[scale=0.62]{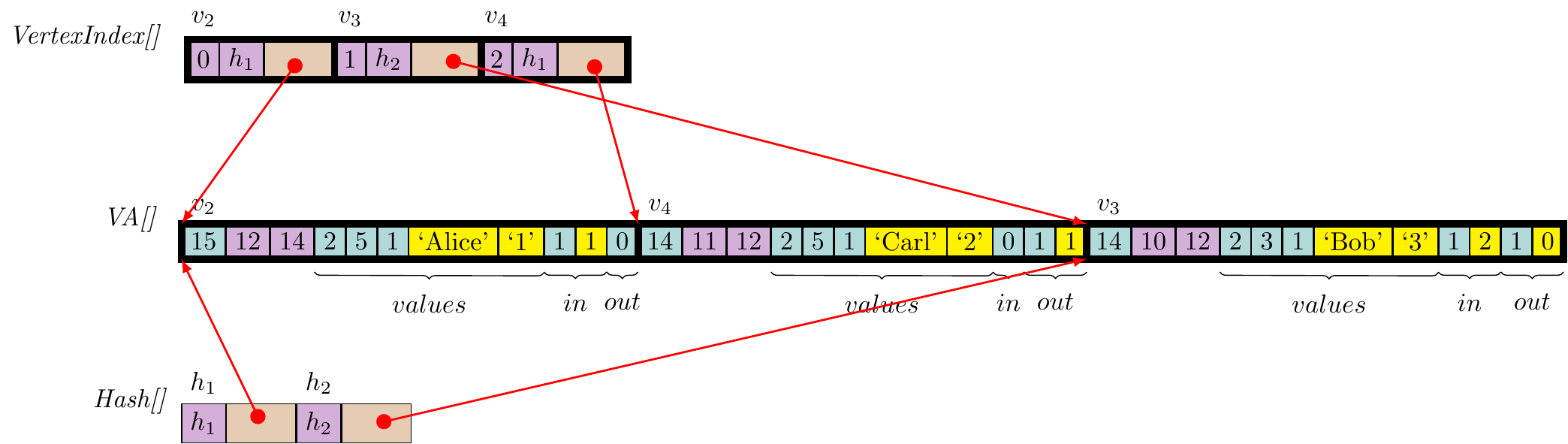}
			\subcaption{Implementation of graph $G_1$ in secondary memory.}\label{fig:graphimplement}
		\end{minipage}%
	\end{adjustbox}
	\caption{\textit{Graph representation in secondary memory}.}
\end{figure*}

\subsection{Implementing the Conjunctive $\theta$-Join\\ through hashing}\label{sec:cogroup}
We propose an implementation of the Conjunctive $\theta$-join
through hashing that uses the CoGrouping \cite{Dittrich}, that is a proposed generalization for
relational tables' join, combining the Sort-Merge Join approach and the Partition-based Hash Join.
With the former approach the tuples are sorted by their join keys, while with the latter
the input is divided in small co-partitions \cite{SchuhCD16}.  In particular, (partition) hash joins
associate all the vertices with a same hash value to a same given bucket and can be only applied over
equi-join predicates \cite{Zeller}. On the other hand, by ordering the buckets by hash value, we can
both decrease the time required to search the buckets with the same hash value, and use the hash
join even when the predicate is a ``$\leq$'' over at most one single attribute per graph, containing ordinal values.
Since we would like to have an algorithm supporting both equijoins and ``$\leq$'' predicates, we use the 
CoGrouped (hash) join technique; hereby we call such algorithm  ``\textbf{CoGrouped Graph Conjunctive $\theta$-Join}''.

As a first step, for each vertex of the left graph $G_1=(V_1,E_1,A)$ we have to reduce the number of
the vertices that have to be scanned in the right graph $G_2=(V_2,E_2,B)$ while searching for a possible match. 
Our aim is to prune the combination of visited pair of distinct vertices, jointly from the left and right graph,
to the sole
matching ones as much as possible. In order to do so
we have to associate to each vertex $u,v$ a value generated from a \textit{vertex hashing function} $h$ such that, if they jointly satisfy the $\theta$ predicate ($\theta(u,v)$), then their values $h(v)$ and $h(v)$
\textit{match}.
If $\theta$ is defined as an equivalence predicate over (a subset of) the attributes in $A$ and $B$  such as:
\begin{equation}
\label{key:qe}
\theta(u,v)=u.A_{i_1}=v.B_{j_1}\wedge\dots\wedge u.A_{i_m}=v.B_{j_m}
\end{equation}
then we have that $h$ has to be defined over the values of $A_{i_1},\dots,A_{i_m}$ and $B_{j_1},\dots,B_{j_m}$, and the hash values \textit{match} iff 
$h(u)=h(v)$. Hereby we shall impose that if  the vertices $u$ and
$v$ satisfy the equi-join predicate in Equation \ref{key:qe}, then such vertices
have the same hash value ($h(u)=h(v)$).

This technique also applies whether the predicate $\theta$ is defined as a total
order over at most one attribute of a vertex (e.g. $\theta(u,v)=u.A\leq v.B$) and in this case the hashing function
should be monotone w.r.t. the ordered values (if $u.A\leq v.B$, then the hashes \textit{match} with $h(u)\leq h(v)$).

In our case study we examine the algorithm when the join predicate is expressed in an
equivalence form such as Equation \ref{key:qe}.
On the other hand, Appendix \ref{app:leq} provides some details on the case of the less-equal ($\leq$) predicate, both
providing the algorithm and the experimental results.

This  paragraph describes the first part of the algorithm, where the join operation is 
performed among the vertices' tables, $V_1$ and $V_2$, as in the standard relational CoGrouping join scenario. 
By sorting the vertices for both graphs by  hash 
value, the vertices having the same hash value are stored contiguously.
As a result, 
we can directly access to the 
vertices with hash code $h_c$ ($u\in V_1$ and $v\in V_2$ having $h(u)=h_c=h(v)$).
In order to visit only the vertices having hash values that are shared between the two indexed graphs,
we preventively evaluate the set $HI$ storing the hash values
shared on both graphs ($HI:=h(V_1)\cap h(V_2)$).
We still have to check if the predicate $\theta$ is satisfied for each pair of such vertices ($\theta(u,v)$) 
as long as the join condition holds (that is $(u\oplus v)[A]=u\wedge (u\oplus v)[B]=v$). If all those tests
are passed, then the joined vertex $u\oplus v$ is the next vertex candidate for the resulting graph.

The last step of the algorithm involves the creation of the edges between the joined vertices, $u\oplus v$, and their neighbours.
Since the neighbours must be elements of $V_1\Join_\theta V_2$, by graph join definition
we choose, among all the possible neighbours, those that have a hash value $h_d$ appearing in both graphs
($h_d\in HI$). In particular even in this case, the vertices have to pass the previously described tests.

Algorithm \ref{alg:cogrouped} implements the CoGrouped Graph Conjunctive EquiJoin 
described idea in pseudocode with no assumptions 
regarding both the operands' and the result's data structure. Better performances can be achieved by
using ad hoc graph data representations, as described in the following section.

\begin{algorithm}[!b]
	\caption{CoGrouped Graph Conjunctive EquiJoin}\label{alg:cogrouped}
	\begin{algorithmic}[1]
		\Procedure{CoGroupedJoin}{$G_1,G_2,\theta$}\Comment{$G_1\Join_\theta^{\wedge} G_2$}
		\State $V\leftarrow \emptyset;\; E\leftarrow \emptyset$
		\State $HI\leftarrow h(V_1)\cap h(V_2)$
		\For{\textbf{each} $h_c\in HI$}
		\For{\textbf{each} $u\in V_1,v\in V_2$ \textbf{s.t.} $h(u)=h_c=h(v)$}
		\If{$\theta(u,v), (u\oplus v)[A]=u, (u\oplus v)[B]=v$}
		\State  $V\leftarrow V\cup \{u\oplus v\}$
		\For{\textbf{each} $nu\in out_{V_1}(u)$}
		\If{$h(nu)\notin HI$} \textbf{continue}
		\EndIf
		\For{\textbf{each} $nv\in out_{V_2}(v)$}
		\If{$h(nv)\neq h(nu)$} \textbf{continue}
		\EndIf
		\If{$\theta(nu,nv),$\par 
			\hskip\algorithmicindent\hskip\algorithmicindent\hskip\algorithmicindent\hskip\algorithmicindent\hskip\algorithmicindent $(nu\oplus nv)[A]=nu,$\par 
			\hskip\algorithmicindent\hskip\algorithmicindent\hskip\algorithmicindent\hskip\algorithmicindent\hskip\algorithmicindent$(nu\oplus nv)[B]=nv$}
		\State {$V\leftarrow V\cup \{nu\oplus nv\}$}
		\State {$E\leftarrow E\cup \{(u\oplus v, nu\oplus nv)\}$}
		\EndIf
		\EndFor
		\EndFor
		\EndIf
		\EndFor
		\EndFor
		\EndProcedure
	\end{algorithmic}
\end{algorithm}

\pagebreak
\section{Graph Data Structures}\label{ref:structure}
In this section we show the two different strategies that our algorithm uses
in order to compute the previously described algorithm efficiently. We first
describe the result graph data structure that has to allow a fast element
insertion and exiguous memory occupation (Section \ref{sub:bulkgraph}), and then
the secondary memory data structure that allows a quick scan of the graph's vertices
(Section \ref{sub:secstor}).

\subsection{Bulk Graph}\label{sub:bulkgraph}
The graph resulting from a join graph query could be too large to fit in 
main memory. For this reason some implementations like RDF4J do not explicitly store 
the result unless explicitly required, but actually evaluate the query step by step.
On the other hand Neo4J sometimes fails to accomplish in providing the final result 
due to the employment of all the available RAM memory (Table \ref{tab:evaluatejoin}). 
In order to allow the storage of the whole result in
main memory we chose to implement an ad-hoc graph data structure represented as
an adjacency list, where each entry represents a result's vertex and where each
vertex's neighbour is only represented by its id. 

\subsection{Secondary Storage}\label{sub:secstor}
The proposed algorithm shows that we can achieve the advantages of
a CoGrouped join iff the vertices are all sorted by hash value.
If such hash-sorted vertices are stored in a linear data structure as in Figure \ref{fig:graphstructure} (e.g. an array, $VA\texttt{[]}$), 
then we could create an hash index (e.g. $Hash\texttt{[]}$) where each record stores the hash value \texttt{value}
and the offset \texttt{VAOffset}
of \textit{VA} where to retrieve the vertices with the same hash. 

Let us now focus on \textit{VA}'s vertices.
Each vertex in \textit{VA} is stored by omitting the vertices' attributes and storing only the values (\texttt{val[1]\dots val[M]})
and, in order to avoid data replication, each incoming vertex \texttt{in[i]} and outgoing vertex \texttt{out[j]} is stored only by its id.

Given that \textit{VA}'s vertices have variable data size, we need
another linear data structure (e.g. $VertexIndex\texttt{[]}$) for accessing efficiently   such vertices. The \texttt{Id} vertex is stored at the  record with number \texttt{Id} in $VertexIndex\texttt{[]}$; in such record \texttt{VAOffset}
points out where the current vertex is stored in \textit{VA}. Consequently, 
this indexing data structure allows accessing each vertex in
$O(1)$ time by its \texttt{Id}. Further details on how
to implement such data structure are given in Appendix \ref{spec:implement}, where 
Algorithm \ref{alg:cogrouped} is rewritten using these data structures.

\begin{example}
	The graph depicted in Figure \ref{fig:figjoing1bis} could be represented in
	Figure \ref{fig:graphimplement}, and hence the graph data model implementation
	can be presented as follows:
	\[G_1=(VertexIndex,VA,Hash,\Set{\texttt{User},\texttt{MsgTime1}})\]
	where the first three arrays refer to the graph implementation previously described 
	(Figure \ref{fig:graphstructure}) and $\Set{\texttt{User},\texttt{MsgTime1}}$ refers
	to the attribute schema associated to the graph.
\end{example}

While modern Relational DBMSs uses variants of the B-Tree data structure to store tuples in main memory,
we decided to use a linear data structure, since the most frequent operation for the Join algorithm is
the linear scan of all the vertices that share the same hashing value. More precisely, a visit of a balanced 
binary search tree with $N$ vertices takes $2N$ (because some vertices are visited two times), while the 
visit of a linear data structure with $N$ records takes exactly $N$. 

We can access to our data structure
using memory mapping techniques: by doing so we delegate the Operating System
to handle which pages have to be either cached or retrieved from secondary memory.
As a consequence, no cache has to be implemented in the graph join code, since the whole process
is completely transparent to the programmer.
The design of the previously described algorithm also permits to reduce the amount of page faults,
since all the vertices with the same hash value are always stored in a contiguous block in \textit{VA}.

Going more  specifically on our implementation, the whole code was written in Java.
In order to perform memory mapping I/O over files greater of \texttt{MAX\_INT} size, we had to use
the \texttt{JNA} library, through which it is possible to directly interact with the \texttt{mmap} system call. 
In this way we can potentially address at least 64TB of data in virtual memory without explicitly allocating  
any data value: as a result we avoid creating objects in Java, thus also reducing the amount of work of
the Java Garbage Collector, since all the comparisons are done over the data's in-memory representation.

\begin{table*}[!p]
	\caption{Benchmarking the time ($\varsigma$) required to load and index the LiveJournal graph join operands in secondary memory}
	\label{tab:storing}
	\centering
	\scalebox{0.8}{
	\begin{tabular}{cc|rrr|rr}
		\toprule
		\multicolumn{2}{c}{\textbf{Operands Size ($|V|$)}} &
		\multicolumn{3}{c}{\textbf{Operands Storing Time} (ms)}  & \multicolumn{2}{c}{\textbf{Speed-up}} \\
		Left & Right & Neo4J ($\varsigma_N$) & RDF4J ($\varsigma_R$) & Our Storage Proposal ($\varsigma'$) & Neo4J ($\sfrac{\varsigma_N}{\varsigma'}$) & RDF4J ($\sfrac{\varsigma_R}{\varsigma'}$)\\
		\midrule
		\begin{filecontents*}{creation.csv}
size,neo,rdf,self,spedA,spedB
$10$,79,283,20,17.30$\times$,14.15$\times$
$10^2$,223,519,27,5.15$\times$,9.22$\times$
$10^3$,706,3\,938,106,2.80$\times$,19.12$\times$
$10^4$,3\,198,35\,598,1093,1.42$\times$,17.00$\times$
$10^5$,26\,336,360\,738,16\,472,1.80$\times$,204.03$\times$
$10^6$,625\,425,4\,423\,577,162\,348,3.60$\times$,26.99$\times$
\end{filecontents*}
\csvreader[late after line=\\]{creation.csv}{}{\csvcoli & \csvcoli & \csvcolii & \csvcoliii & \csvcoliv & \csvcolv  & \csvcolvi}
		
		\bottomrule
	\end{tabular}}
\end{table*} \begin{table*}[!p]
\caption{LiveJournal graph Join Running Time ($\tau$). GDBMS are tested with different languages: Neo4J is tested with Cypher and RDF4J is tested with SPARQL}\label{tab:evaluatejoin}
\begin{adjustbox}{max width=\textwidth}
	\begin{tabular}{@{}cc|rl|rrr|rr@{}}
		\toprule
		\multicolumn{2}{c}{\textbf{Operands Size ($|V|$)}} &
		\multicolumn{2}{c}{\textbf{Result}} & \multicolumn{3}{c}{\textbf{Join Time} $\tau$(ms)}  & \multicolumn{2}{c}{\textbf{Speed-up}} \\
		Left & Right & Size ($|V|$) & Avg. Multiplicity & Neo4J-Cypher ($\tau_N$) & RDF4J-SPARQL ($\tau_R$) & Our CoGrouped Join ($\tau'$) & Cypher ($\sfrac{\tau_N}{\tau'}$) & SPARQL ($\sfrac{\tau_R}{\tau'}$) \\
		\midrule
		\begin{filecontents*}{calc2.csv}
"size","sizer","avg","neo4j","rdf4j","proposed","spedA","spedB"
10,4,1.00,187.34,5.78,0.44,425.77$\times$,13.14$\times$
$10^2$,59,1.00,233.08,148.46,4.13,56.44$\times$,35.95$\times$
$10^3$,534,1.00,481.02,25\,393.86,11.29,42.61$\times$,2\,249.23$\times$
$10^4$,5\,466,1.03,2\,937.55,2\,947\,822.35,47.59,61.73$\times$,61\,942.05$\times$
$10^5$,87\,377,1.30,197\,872.78,$>$4H,433.15,456.82$\times$,---
$10^6$,4\,025\,040,4.07,$>$4H,$>$4H,19\,279.58,---,---
\end{filecontents*}
\csvreader[late after line=\\]{calc2.csv}{}{\csvcoli & \csvcoli & \csvcolii & \csvcoliii & \csvcoliv & \csvcolv  & \csvcolvi& \csvcolvii & \csvcolviii}%
		\bottomrule
	\end{tabular}
\end{adjustbox}
\end{table*} \begin{table*}[!p]
\caption{This table sums up the results achieved in the two previous tables. It shows that the sum of our proposal's secondary memory operand store and indexing time plus the graph join time ($\varsigma'+\tau'$) is less than the sole graph join time of with the GDBMSs' query languages ($\tau_N$, $\tau_R$). This motivates the usage of our proposal as an implementation of the join operation over the already existing GDBMSs.}\label{tab:outperf}
\centering
\scalebox{0.8}{
\begin{tabular}{cc|rr|r|rr}
	\toprule
	\multicolumn{2}{c}{\textbf{Operands Size ($|V|$)}} &
	\multicolumn{2}{c}{\textbf{Join Time}  (ms)}  & \multirow{2}{*}{\parbox{5.9cm}{Our proposal's \textbf{storing-indexing graph}  $+$  \textbf{CoGrouped join  time} (ms) ($\varsigma'+\tau'$)}} & \multicolumn{2}{c}{\textbf{Speed-up}} \\
	Left & Right & Neo4J ($\tau_N$) & RDF4J ($\tau_R$) &  & Neo4J ($\frac{\tau_N}{\varsigma'+\tau'}$) & RDF4J ($\frac{\tau_R}{\varsigma'+\tau'}$) \\
	\midrule
	\begin{filecontents*}{sum2.csv}
"size","neo4j","rdf4j","proposed","spedA","spedB"
10,187.34,5.78,20.44,9.17$\times$,0.28$\times$
$10^2$,233.08,148.46,31.13,7.49$\times$,4.77$\times$
$10^3$,481.02,25\,393.86,117.29,4.1$\times$,216.5$\times$
$10^4$,2\,937.55,2\,947\,822.35,1\,140.59,2.58$\times$,2\,584.47$\times$
$10^5$,197\,872.78,$>$4H,16\,905.15,11.7$\times$,—
$10^6$,$>$4H,$>$4H,181\,627.58,---,---
\end{filecontents*}
\csvreader[late after line=\\]{sum2.csv}{}{\csvcoli & \csvcoli & \csvcolii & \csvcoliii & \csvcoliv & \csvcolv  & \csvcolvi}%
	\bottomrule
\end{tabular}}
\end{table*}\begin{table*}[!p]
\caption{LiveJournal join operands occupation in secondary memory expressed in Block Size (\texttt{du -hsk}) on a HFS file system.}
\label{tab:size}
\centering
\scalebox{0.8}{
\begin{tabular}{cc|rrr|rr}
	\toprule
	\multicolumn{2}{c}{\textbf{Operands Size ($|V|$)}} &
	\multicolumn{3}{c}{\textbf{Block Size} (KB)}  & \multicolumn{2}{c}{\textbf{Gain}} \\
	Left & Right & Neo4J ($s_N$) & RDF4J ($s_R$) & Our Storage Proposal ($s'$) & Neo4J ($\sfrac{s_N}{s'}$) & RDF4J ($\sfrac{s_R}{s'}$)\\
	\midrule
	\begin{filecontents*}{store_R.csv}
size,neo,rdf,self,spedA,spedB
$10$,308,76,24,12.83$\times$,3.16$\times$
$10^2$,408,204,32,12.75$\times$,6.37$\times$
$10^3$,1\,608,1\,616,216,7.44$\times$,7.48$\times$
$10^4$,13\,740,16\,124,6.53$\times$,7.66$\times$
$10^5$,141\,540,153\,876,20\,660,6.85$\times$,7.45$\times$
$10^6$,1\,458\,868,1\,438\,748,193\,772,7.53$\times$,7.42$\times$
\end{filecontents*}
\csvreader[late after line=\\]{store_R.csv}{}{\csvcoli & \csvcoli & \csvcolii & \csvcoliii & \csvcoliv & \csvcolv  & \csvcolvi}%
	\bottomrule
\end{tabular}}
\end{table*} \begin{figure*}[!p]
	\centering
\begin{adjustbox}{max width=.9\textwidth}
	\hspace*{-0.5cm}  
	\centering
 \begin{minipage}[b]{0.6\textwidth}
	\hspace{1cm}\includegraphics[scale=0.35]{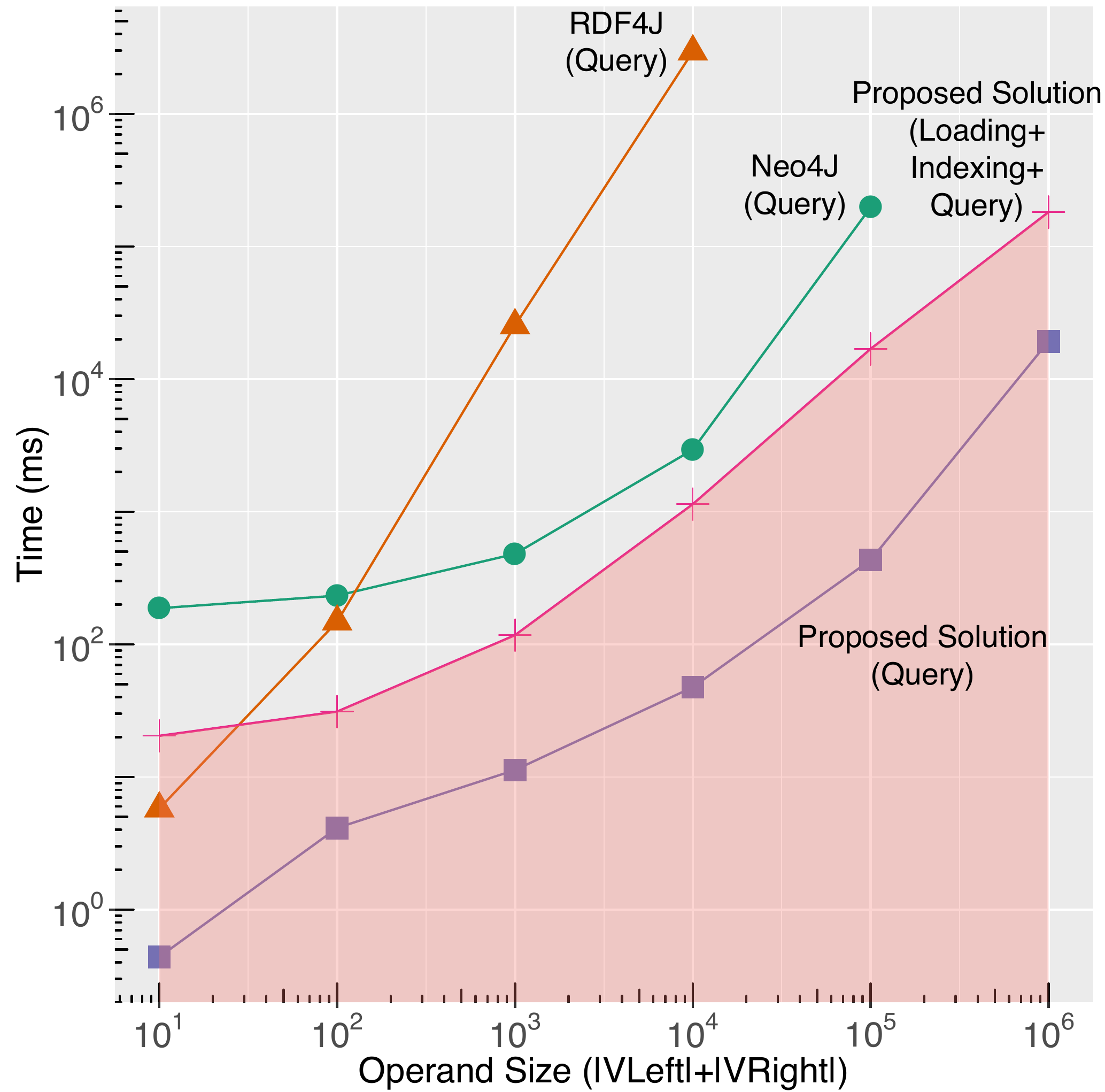}
	\subcaption{Comparing current solutions' query time with our proposal's loading+indexing+query time for the LiveJournal Social Network Graph (Table \ref{tab:outperf}).}\label{fig:sumsize}
\end{minipage} \qquad
	\begin{minipage}[b]{.5\textwidth}
		\includegraphics[scale=0.35]{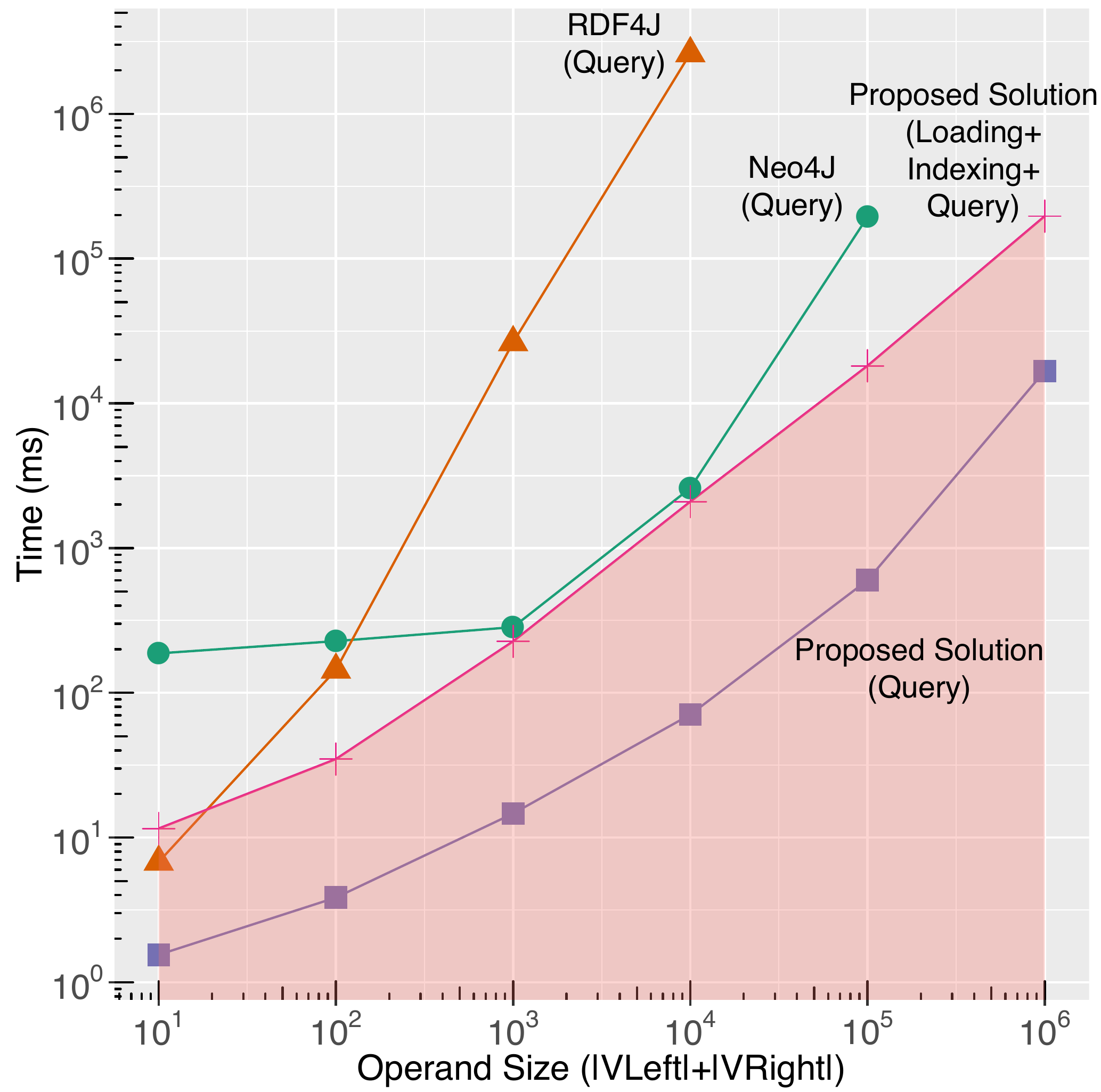}
		\subcaption{Comparing current solutions' query time with our proposal's loading+indexing+query time for the YouTube Graph.}\label{fig:sumsizey}
	\end{minipage}
\end{adjustbox}
\caption{\textit{Plots for the benchmark tests}.}
\end{figure*}

\section{Dataset}\label{sec:dataset}
We tested our algorithm using the LiveJournal  \cite{dataSIOC} and the 
YouTube \cite{Yang2015} Social Network Graphs. The former
contains  4,847,571 unlabelled vertices and 68,993,773 edges, while the latter contains 
1,134,890 unlabelled vertices and 2,987,624 edges. In both graphs each vertex represents a 
user which is connected to one if its friends by an edge. For both graphs we followed the same
procedure for obtaining our graph join operands through random walk sampling. 
Such procedure is described below.

Since no data values are given within the datasets, we
enriched each graph: we used the guidelines of the 
LDBC Social Network Benchmark protocol \cite{Erling}, and hence associated to each user an
IP address, an Organization and the year of employment\footnote{The resulting enriched graph
	is available at \url{http://jackbergus.alwaysdata.net/BolognaGraph2016.tar.gz}}. We choose to not use the whole LDBC
model, both because we prefer to test our dataset with real world graph data, and because those 
guidelines are specifically built for generating RDF graphs. 

For each social network graoh, the left and right graph operands were obtained by starting the random walk from a same vertex
but using a different seed for the graph traversal. Moreover, such operands were obtained incrementally
by visiting each time a number of vertices that is a power of $10$. The operands were stored in secondary memory 
as Neo4J graphs, as RDF4J native triple stores and with our proposed implementation. Since
Neo4J's query language (Cypher) does not support querying two distinct graphs, both left and
right operands were stored in the same graph database.

Before discussing the  indexing implementation choices for Neo4J, RDF4J and our proposed implementation,
we must discuss the graph join operation that will be benchmarked.
We perform the algorithm using as operands the two distinct subgraphs with a same vertex size, where the 
$\theta$ predicate is the following one:\[\begin{split}
\theta(u,v)\overset{def}{=}&u.Year1 = v.Year2\; \wedge\\ 
& u.Organization1 = v.Organization2
\end{split}\]
In this way we generalize the ``subgraph extraxtion from two graphs'' problem by matching not only
one user per time, but also all the users with the same employment year and  work for the same organization. The resulting graph has
an edge between two vertices iff there was a friendship relation among the two users in the 
original graph.

In order to compare our indexed implementation with other indexing features of both
graph databases, for Neo4J we create the indices over the $Year$ and $Organization$ attributes and,
for the RDF triple store where no properties are associated to the vertices, we
create all the indices over the stored triples (subject, object and predicate).
While Neo4J can directly store the data as described by our data model, 
RDF4J requires to translate such data since it uses the RDF data model.
In order to do so, for each attribute $A_i$ of our vertex $v$
we create a triple $(v,A_i,v[A_i])$, representing two nodes, $v$ and $v[A_i]$, linked by 
an edge $A_i$.  
Regarding our proposed data structure, we associate to each vertex the hashing value 
obtained from an automatically generated hashing composition between the hashes of the
field $Year$ and $Organization$. The hash function used for both fields are Java standard
hashing functions.

\section{Results}\label{ref:results}

We evaluated our proposed implementation against Neo4J and RDF4J with a computer with
a 2.2 GHz Intel Core i7 processor and 16 GB of RAM at 1600 MHz, and an SSD Secondary Storage with an HFS file system. Moreover, Neo4J is already optimized for SSD \footnote{``I presume you use a sensible machine, with a SSD (or enough IOPS) and decent amount of RAM.''
	\texttt{\small http://neo4j.com/blog/neo4j-2-2-query-tuning/}}, and all triple stores have been 
tested with SSD secondary storage\footnote{RDF4J is previously known as Sesame: \texttt{\small https://www.w3.org/wiki/LargeTripleStores}}. All the queries (either in Cypher or SPARQL) have
been evaluated using Java: this means that our solution has been implemented in Java\footnote{See our code in \url{https://bitbucket.org/unibogb/databasemappings/}, released under GPLv3.}
using the JNA library for calling the OS memory mapping methods, and the Java APIs for
both Neo4J and RDF4J. We chose those two graph databases because they
implement two different graph query languages that are based upon distinct graph
data model: Neo4J stores Property Graphs and queries the graphs with Cypher,
while RDF4J is a triple store that queries the graph data with SPARQL. We chose 
not to test our graph joins over the SQLGraph model \cite{SQLGraph} since there 
is no existing implementation of such model and, most importantly, the Gremlin query language allows
only to perform graph traversal queries that usually return a bag of values
(see Section \ref{sec:dbqlang} for further details).

While Neo4J explicitly produces the result in the graph
database and returns it through the API interface, RDF4J explicitly evaluates 
the query either by creating a new graph in main memory or by iterating over an
API object. Consequently, we choose to evaluate RDF4J while storing the result
in a main-memory graph. 

All the benchmarks
are produced by performing a trimmed mean of $10$ runs over
the same dataset. In such trimmed mean 
the fastest and the slowest results are removed. 
Since Cypher stores the newly created vertices and edges
into the secondary memory operands' property graphs, after each run we have
to completely delete the database in order to perform with no result stored in the graph
database. 
As a consequence we decided to sequentially perform the benchmarks
over the different graph data sizes for each GDBMS  (i.e. we first perform the join over the
$10\times 10$ operands, then over the $10^2\times 10^2$ operands and so on), 
and to perform the next run for Neo4J after re-creating the graph database in secondary
memory. For the next run we recreate all the graph databases, and we recommence
the mentioned process. Our solution outperforms the implementation over Neo4J  an order 
of magnitude and at most two orders of magnitude, while RDF4J performances degrades 
drastically when reaching graph operands with larger size (we outperform by at most
five orders of magnitude with $10^4$ vertices' operand size). Moreover, we set a timeout of
four hours (4H) while performing the benchmark tests.

Table \ref{tab:evaluatejoin} provides the time that is required 
to evaluate the join operation for the LiveJounral dataset. The operand size is expressed with the number of the graphs' 
vertices. We can see that our solution outperforms 
the two benchmarked systems: while RDF4J behaves better than Neo4J on data with small 
operand size ($10$, $10^2$), the opposite situation happens when operands have greater size
(up to $10^6$). A graphical representation is given in  Figure \ref{fig:sumsize}.

In particular, Figure \ref{fig:sumsizey} plots the same results for the YouTube social network
graphs: as we can see, even if the speed-up in such solution is lower than in the previous
dataset, our join implementation still outperforms the opponents' graph join implementations.

Table \ref{tab:storing} shows that our implementation is faster to store than both 
Neo4J's property graph and RDF4J's triple store. 
Moreover, our graph data structure
requires less secondary memory blocks to be stored (Table \ref{tab:size}). The constant
gain of our implementation against Neo4J is supported by the fact that Neo4J
has a secondary memory cache in order to speed up the graph traversal query process.
On the other hand, RDF4J through the Sail storage represents the RDF triples into a 
B-Tree structure.
Anyway, we must clarify that our data structures contain neither locks nor cache areas for other
graph operations, that are usually implemented in all-round GDBMS solutions.

Table \ref{tab:outperf} 
shows that the time
required for both loading and indexing the operands in secondary memory and join them
is still lesser than the sole join query time for Neo4J and RDF4J
(except for the case of the operand size $10$ over RDF4J, when RDF4J outperforms our
proposed solution).

\section{Conclusions}\label{src:conc}
We propose a new graph operation that, to the best of our knowledge,
is not supported on current graph query languages, that is the graph join. This is a binary
operation performing a relational join over the graphs' vertices tables that, in a second step, creates 
the edges according to a specific semantics. 

The results highlight how our proposed graph data structure and join algorithm can be 
used for implementing the graph join operation over the state-of-the-art graph database 
management systems. This is due to the fact that the time that it takes to both store
the proposed graph data structure in secondary memory  and to join them  is less than the
sole query time of such systems ($\tau_N>\varsigma'+\tau'$ and $\tau_R>\varsigma'+\tau'$). 
Our graph join algorithm assumed that the 
graph operand's edges are not labelled: further tests have to be carried out to both 
extend our data structure and our algorithm, in order to allow more of one type of edges at a time.

\subsection{Future Works}
The array data structure of our graph data model allows implementing the join
algorithm with a parallel approach: we could easily assign to each process a subset of 
vertices with a specific hash code, store the result into a subset bulk graph and
to append the partial results, by linking the end of one stack with the beginning of
the next partial computation.

As a next step we could see that it is possible to define a specific case of $\theta$-join when the
relation $\theta$ is defined over some edges that can link either two distinct graphs, or the edges of
a same given graph. Consequently, given a set of edges $E$, $\Re_E$ is defined as the relation
$v\Re_E v'\Leftrightarrow (v,v')\in E$. A specific application of such kind of join can be found in OrientDB's (No)SQL language,
where no joins, either explicit or implicit in the from clause (multiple tables) are 
allowed between vertices.

{
	The Disjunctive Join acts as a missing 
	data operator for the edges.
	The disjunctive join does not allow inheriting missing information from the two graph operands, since
	only the vertices that match over the two graphs are returned.
	Consequently, as done in the relational data model,
	we could provide more details on graph outer joins (such as left, right and full). While the graph conjunctive join acts as an intersection between the
	two graphs, the graph disjunctive full-outer join could act as a graph union operation. 
}

\section{Related Work}\label{sec:dbqlang}
Graph Query Languages can be categorised in two main classes; the first class of languages try to find a
possible match for a specific traversal expression (Section \ref{subsec:traversal}): as a consequence such graph 
queries do not manipulate the graph data structure. The second class of queries use traversal expressions to extract
a portion of the graph that has to be queried and altered in a next step.

\subsection{Graph Traversal lanugages}\label{subsec:traversal}

\phparagraph{GraphLOG}
The \textbf{GraphLOG} \cite{graphlog} query language subsumes a property graph 
data structure (\textit{direct labelled multigraph})  where no properties are associated, neither to vertices 
nor to edges. Such query language is conceived to be visually representable, and hence path queries are
represented as graphs, where simple regular expressions can be associated to the edges. The concept of
visually representing graph traversal queries involving path regex-es was later on adopted in \cite{n3},
where some algorithms are showed for implementing such query language in polynomial time. Such language
does not support some path summarization queries that were introduced in GraphLOG \cite{GraphLogAggr}. 

\phparagraph{NautiLOD}
The \textbf{NautiLOD} \cite{NautiLOD} query language was conceived for performing path queries (defined through
regular expressions) over 
RDF graphs with recursion operators (Kleene Star). The same paper shows  that queries can be evaluated in polynomial time.

\phparagraph{Gremlin}
Another graph traversal language, \textbf{Gremlin}, have been proved to be Turing Complete \cite{Rodriguez15}:
by the way this is not a desired feature for query languages since it must guarantee that 
each query always returns an answer and that the evaluation of the query always terminates.
Another problem with this query language is based on its semantics:
while all the other graph traversal languages return the desired subgraph, Gremlin returns
a bag of values (e.g. vertices, values, edges). This peculiarity does not allow the user to take advantage 
of partial query evaluations and to combine them in a final result.

\subsection{(Proper) Graph Query Languages}\label{subsec:proper}
All the following graph query languages offer only a limited support to pattern extraction from graphs,
except from SPARQL, that has been recently extended in order to allow path traversal queries \cite{Kostylev2015}.
Consequently, such languages focus more on the graph data manipulation part.


\phparagraph{BiQL}
\textbf{BiQL} \cite{BiQL,BiQL2} is a SQL-like query language that allows to (i) update the data graph with new vertices
and edges, (ii) filter the desired vertices through a \texttt{WHERE} clause, (iii) extract desidered subgraph through 
path expressions and (iv) provide some basics path and vertex summarization results. This language has not got a 
formal semantics yet and it is still under development, but has the aim to develop a closed language under query
compositionality. The query patterns do not allow expressing regex-es over the paths.

\phparagraph{SPARQL}
At the time of writing, the most studied graph query language that has been studied both in terms of
semantics and expressive power is \textbf{SPARQL}, as it is the most time-worn language among those that are both well-known
and implemented. Some studies on the expressive power of SPARQL \cite{SparQLExpr,SPARQLSem} showed that
such graph query language syntax allows to write very costly queries that can be computed more 
efficiently whether only a specific class of (equivalent) queries is allowed. As a result, the design flaws
of a query language relapse on the computational cost of the allowed queries. These problems could be avoided 
from the very beginning whether the formal study had preceded the practical implementation of the language.
However, such limitations do not preclude some interesting properties: the algebraic language used to
formally represent SPARQL performs queries' incremental evaluations \cite{SparqlIncr}, and hence
allows to boost the querying process while data undergoes updates (both incremental and decremental). 
Moreover, while SPARQL was originally designed to return tabular results, 
later extensions tried to overcome to such problem with the \texttt{CONSTRUCT} clause,
that returns a new graph. Last but not least, the usage of so-called \textit{named graphs}
allows to perform queries over two distinct RDF graphs.

\phparagraph{LDQL}
The NautiLOD language was later on extended in \textbf{LDQL} \cite{Hartig2015,HartigP15a}, where SPARQL patterns are added
and different path union and concatenation are allowed. For this specific graph query language the time
complexity of the query evaluation has not been studied in the time of the writing. Even if it is claimed that 
such language is more general than SPARQL, this language do not allow to create new graphs and to concatenate
vertex values through the \texttt{BIND} clause, since most of the SPARQL operations are not matched
by the sole RDFs triple matching.

\phparagraph{Cypher}
Cypher \cite{Neo4jMan,Robinson,CypherCheat} is yet another SQL-like graph query language for property graphs. No formal semantics for this
language were defined from the beginning as in GraphQL, but nervelessly some theoretic results have been carried out 
for a subset of \textbf{Cypher} path queries \cite{Neo4jAlg} by using an algebra adopting a path implementation
over the relational data model. Similarly to BiQL, this language does not allow to express complex
graph traversal patterns, but it allows to update a property graph and to produce a new graph
as a result.

\phparagraph{GraphQL and GRAD}
\textbf{GraphQL} \cite{HePhD} is yet another graph query language with an SPARQL-like syntax, mainly conceived
for pattern extraction from the data, called \textit{graph motifs}, and their construction. The language
allows graphs naming similarly to SPARQL \textit{named graphs}. The most interesting scientific contribution
of He \cite{HePhD} is the first attempt in defining a graph algebra for collection of graphs. This approach
has been finally specialized for single graphs in the \textbf{GRAD} algebra \cite{GRAD,Ghrab2015}. In this
latter definition the \textit{cartesian product} and \textit{join} operations are still defined over 
graph collections and are still not specialized for the single graphs. Consequently, in both languages
the cartesian product over two graph collections produces a graph containing two (possibly) disjoint graph
components. The graph join over the two collections only merges the matched vertices and no considerations
are made on the graphs' edges structure. 

In the end, GRAD propose an alternative
graph data model that could be expressed as a specific implementation of the Property Graph model. 

\subsection{Proposed Graph Products and Joins}
\phparagraph{Discrete Mathematics}
At the time of writing, the only field where graph joins where effectively 
discussed is Discrete Mathematics. In this field such operations are defined 
over either on finite graphs or on finite graphs with cycles, and are named
\textit{graph products} \cite{Hammack}. As the name suggests, every graph product of two graphs,
e.g. $G_1=(V_1,E_1)$ and $G_2=(V_2,E_2)$, produces a graph whose vertex set 
is defined as $V_1\times V_2$, while the edge set changes accordingly to the
different graph product definition. Consequently the Kroneker Graph Product
\cite{Weichsel} is defined as follows:
\begin{gather*}
G_1\times G_2=(V_1\times V_2, \Set{((g,h),(g',h'))\in V_1\times V_2|(g,g')\in E_1,(h,h')\in E_2})
\end{gather*}
while the \textit{cartesian graph product} \cite{ImrichP07} is defined as follows:
\begin{gather*}
G_1\square G_2=(V_1\times V_2,\Set{((g,h),(g',h'))\in V_1\times V_2|(g=g',(h,h')\in E_2) \vee  (h=h', (g,g')\in E_1)})
\end{gather*}
Please observe that this definition creates a new vertex which is a pair of 
vertices: hereby such operation is defined differently from the relational
algebra's cartesian product, where the two vertices are merged. As a consequence,
such graph products admit commutativity and associativity properties only up to
graph isomorphism. Other graph products are \textit{lexicographic product}
and \textit{strong product} \cite{Hammack,ProductGraphs}.

\begin{table*}[!t]
	\centering
	\begin{adjustbox}{max width=\textwidth}
		\begin{threeparttable}
			\begin{tabular}{|c|c|c|c|c|c|c|c|} \hline
				\small
				& \multicolumn{2}{c}{Closure Property}\vline & \multicolumn{4}{c}{Graph Operators} \vline& \\
				& Input\tnote{a} & Output\tnote{a} & \shortstack{Update \\ Values} & Summarization\tnote{a} & Graph Join & \shortstack{Traversal with \\ Branches} & \shortstack{Incremental \\ Update} \\\hline
				
				GraphLOG & Labelled Graph 
				& Labelled Graph
				& \xmark 
				& \cmark(over Paths)
				& \xmark 
				& \cmark, graphical, RegEx
				& \xmark\\\hline

				NautiLOD & WLOD 
				& WLOD 
				& \xmark 
				& \xmark
				& \xmark 
				& \cmark, disjunctive, RegEx
				& \xmark\\\hline
				
				Gremlin & PG
				&  \shortstack{Bag of vertices,\\ edges or values} 
				& \xmark 
				& \cmark (Bag)  
				& \xmark 
				& \cmark 
				& Titan Script\tnote{b}\\\hline
				
				
				BiQL & PG, only edge weight
				& PG, only edge weight
				& \cmark
				& \cmark (PG)
				& \xmark
				& \xmark
				& \xmark\\\hline
				
				SPARQL & RDF Graph 
				& Table, Graph
				& \xmark 
				& \xmark 
				& \xmark 
				& \shortstack{Triples with variables,\\ Property Path} 
				& \shortstack{With \cite{SparqlIncr} \\ algebra}\\\hline
				
				LDQL & WLOD 
				& WLOD
				& \xmark 
				& \xmark 
				& \xmark 
				& \shortstack{SPARQL Triples,\\ NautiLOD} 
				& \xmark \\\hline
				
				Cypher & PG 
				& PG or Tables 
				& \cmark
				& \cmark (Tables)
				& \xmark
				&  \shortstack{\cmark, no RegEx} 
				& \xmark\\\hline
				
				GRAD & GRAD 
				& GRAD 
				& \xmark
				& \xmark 
				& \xmark
				&  \shortstack{\xmark, separately in GraphQL} 
				& (possible) \\\hline
				
			\end{tabular}
			\begin{tablenotes}
				\small
				\item[a] PG, shorthand for Property Graph
				\item[b] \texttt{http://s3.thinkaurelius.com/docs/titan/0.5.0/hadoop-distributed-computing.html}
			\end{tablenotes}
		\end{threeparttable}
	\end{adjustbox}
	\caption{Comparing graph query languages}
	\label{tab:summinglang}
\end{table*}
\phparagraph{Graph Database}
Table \ref{tab:summinglang} summarizes some graph database
languages features that were previously described in Section \ref{sec:dbqlang}.
As we can see, no previously described graph query language has a join operand
between graphs. Consequently, our paper proposes such operand and outlines an
algorithm that can be used in order to implement some cases of the graph
conjunctive join. As previously described in the introduction, 
the term ``join'' in graph databases has been used to indicate 
other types of non-binary graph operations: some papers 
\cite{LiM03,Holzschuher,Gao,Zou09} use the term ``path join'' as 
path queries of arbitrary length, while others more specifically
\cite{Atre,Yuan,Fletcher09} refer to joins between adjacent vertices.
While the first definition mainly focuses
on path extraction from a single graph as a specific 
instance of a pattern extraction problem, the second 
problem focuses on the definition of new edges as a 
result of graph traversal queries. 

\pagebreak
\bibliographystyle{abbrv}
\bibliography{sigproc}

\begin{thebibliography}{10}

\bibitem{SparQLExpr}
R.~Angles and C.~Gutierrez.
\newblock {\em The Semantic Web - ISWC 2008: 7th International Semantic Web
  Conference, ISWC 2008, Karlsruhe, Germany, October 26-30, 2008. Proceedings},
  chapter The Expressive Power of SPARQL, pages 114--129.
\newblock Springer Berlin Heidelberg, Berlin, Heidelberg, 2008.

\bibitem{SIGMOD2015Atre}
M.~Atre.
\newblock {Left Bit Right: For SPARQL Join Queries with OPTIONAL Patterns
  (Left-outer-joins)}.
\newblock In {\em {SIGMOD Conference}}, pages 1793--1808. {ACM}, 2015.

\bibitem{Atre}
M.~Atre, V.~Chaoji, M.~J. Zaki, and J.~A. Hendler.
\newblock Matrix "bit" loaded: A scalable lightweight join query processor for
  rdf data.
\newblock In {\em Proceedings of the 19th International Conference on World
  Wide Web}, WWW '10, pages 41--50, New York, NY, USA, 2010. ACM.

\bibitem{Berlingerio11}
M.~Berlingerio, M.~Coscia, and F.~Giannotti.
\newblock Finding redundant and complementary communities in multidimensional
  networks.
\newblock In {\em Proceedings of the 20th ACM International Conference on
  Information and Knowledge Management}, CIKM '11, pages 2181--2184, New York,
  NY, USA, 2011. ACM.

\bibitem{Boden12}
B.~Boden, S.~G\"{u}nnemann, H.~Hoffmann, and T.~Seidl.
\newblock Mining coherent subgraphs in multi-layer graphs with edge labels.
\newblock In {\em Proceedings of the 18th ACM SIGKDD International Conference
  on Knowledge Discovery and Data Mining}, KDD '12, pages 1258--1266, New York,
  NY, USA, 2012. ACM.

\bibitem{preSQLGraph}
M.~A. Bornea, J.~Dolby, A.~Kementsietsidis, K.~Srinivas, P.~Dantressangle,
  O.~Udrea, and B.~Bhattacharjee.
\newblock Building an efficient rdf store over a relational database.
\newblock In {\em Proceedings of the 2013 ACM SIGMOD International Conference
  on Management of Data}, SIGMOD '13, pages 121--132, New York, NY, USA, 2013.
  ACM.

\bibitem{GraphLogAggr}
M.~Consens and A.~Mendelzon.
\newblock Low complexity aggregation in graphlog and datalog.
\newblock In S.~Abiteboul and P.~Kanellakis, editors, {\em ICDT '90}, volume
  470 of {\em Lecture Notes in Computer Science}, pages 379--394. Springer
  Berlin Heidelberg, 1990.

\bibitem{graphlog}
M.~P. Consens and A.~O. Mendelzon.
\newblock Graphlog: A visual formalism for real life recursion.
\newblock In {\em Proceedings of the Ninth ACM SIGACT-SIGMOD-SIGART Symposium
  on Principles of Database Systems}, PODS '90, pages 404--416, New York, NY,
  USA, 1990. ACM.

\bibitem{Dittrich}
J.~Dittrich.
\newblock {\em Patterns in Data Management: A Flipped textbook}.
\newblock Jens Dittrich, Saarland University, Germany, 1 edition, 2016.

\bibitem{BiQL}
A.~Dries, S.~Nijssen, and L.~De~Raedt.
\newblock A query language for analyzing networks.
\newblock In {\em Proceedings of the 18th ACM Conference on Information and
  Knowledge Management}, CIKM '09, pages 485--494, New York, NY, USA, 2009.
  ACM.

\bibitem{BiQL2}
A.~Dries, S.~Nijssen, and L.~Raedt.
\newblock {\em Bisociative Knowledge Discovery: An Introduction to Concept,
  Algorithms, Tools, and Applications}, chapter BiQL: A Query Language for
  Analyzing Information Networks, pages 147--165.
\newblock Springer Berlin Heidelberg, Berlin, Heidelberg, 2012.

\bibitem{Erling}
O.~Erling, A.~Averbuch, J.~Larriba-Pey, H.~Chafi, A.~Gubichev, A.~Prat, M.-D.
  Pham, and P.~Boncz.
\newblock The ldbc social network benchmark: Interactive workload.
\newblock In {\em Proceedings of the 2015 ACM SIGMOD International Conference
  on Management of Data}, SIGMOD '15, pages 619--630, New York, NY, USA, 2015.
  ACM.

\bibitem{n3}
W.~Fan, J.~Li, S.~Ma, N.~Tang, and Y.~Wu.
\newblock Adding regular expressions to graph reachability and pattern queries.
\newblock {\em Frontiers of Computer Science}, 6(3):313--338, 2012.

\bibitem{NautiLOD}
V.~Fionda, G.~Pirrò, and C.~Gutierrez.
\newblock Nautilod: A formal language for the web of data graph.
\newblock {\em TWEB}, 9(1):5:1--5:43, 2015.

\bibitem{Fletcher09}
G.~H. Fletcher and P.~W. Beck.
\newblock Scalable indexing of rdf graphs for efficient join processing.
\newblock In {\em Proceedings of the 18th ACM Conference on Information and
  Knowledge Management}, CIKM '09, pages 1513--1516, New York, NY, USA, 2009.
  ACM.

\bibitem{Gao}
J.~Gao, J.~Yu, H.~Qiu, X.~Jiang, T.~Wang, and D.~Yang.
\newblock Holistic top-k simple shortest path join in graphs.
\newblock {\em IEEE Trans. on Knowl. and Data Eng.}, 24(4):665--677, Apr. 2012.

\bibitem{Ghrab2015}
A.~Ghrab, O.~Romero, S.~Skhiri, A.~Vaisman, and E.~Zim{\'a}nyi.
\newblock {\em Advances in Databases and Information Systems: 19th East
  European Conference, ADBIS 2015, Poitiers, France, September 8-11, 2015,
  Proceedings}, chapter A Framework for Building OLAP Cubes on Graphs, pages
  92--105.
\newblock Springer International Publishing, Cham, 2015.

\bibitem{GRAD}
A.~Ghrab, O.~Romero, S.~Skhiri, A.~A. Vaisman, and E.~Zimányi.
\newblock Grad: On graph database modeling.
\newblock {\em CoRR}, abs/1602.00503, 2016.

\bibitem{Hammack}
R.~Hammack, W.~Imrich, and S.~Klavzar.
\newblock {\em Handbook of Product Graphs, Second Edition}.
\newblock CRC Press, Inc., Boca Raton, FL, USA, 2nd edition, 2011.

\bibitem{Hartig2015}
O.~Hartig and J.~P{\'e}rez.
\newblock {\em The Semantic Web - ISWC 2015: 14th International Semantic Web
  Conference, Bethlehem, PA, USA, October 11-15, 2015, Proceedings, Part I},
  chapter LDQL: A Query Language for the Web of Linked Data, pages 73--91.
\newblock Springer International Publishing, Cham, 2015.

\bibitem{HartigP15a}
O.~Hartig and J.~Pérez.
\newblock Ldql: A query language for the web of linked data (extended version).
\newblock {\em CoRR}, abs/1507.04614, 2015.

\bibitem{HePhD}
H.~He.
\newblock {\em Querying and Mining Graph Databases}.
\newblock PhD thesis, Santa Barbara, CA, USA, 2007.
\newblock AAI3283657.

\bibitem{Holzschuher}
F.~Holzschuher and R.~Peinl.
\newblock Querying a graph database language selection and performance
  considerations.
\newblock {\em Journal of Computer and System Sciences}, 82(1, Part A):45 --
  68, 2016.
\newblock Special Issue on Query Answering on Graph-Structured Data.

\bibitem{Neo4jAlg}
J.~Hölsch and M.~Grossniklaus.
\newblock An algebra and equivalences to transform graph patterns in neo4j.
\newblock {\em Fifth International Workshop on Querying Graph Structured Data},
  2016.

\bibitem{ProductGraphs}
W.~Imrich and S.~Klavzar.
\newblock {\em Product Graphs. Structure and Recognition}.
\newblock John Wiley \& Sons, Inc., New York, NY, USA, 2nd edition, 2000.

\bibitem{ImrichP07}
W.~Imrich and I.~Peterin.
\newblock Recognizing cartesian products in linear time.
\newblock {\em Discrete Mathematics}, 307(3-5):472--483, 2007.

\bibitem{CypherCheat}
N.~T. Inc.
\newblock Cypher cheat scheet, 2014.

\bibitem{Kostylev2015}
E.~V. Kostylev, J.~L. Reutter, M.~Romero, and D.~Vrgo{\v{c}}.
\newblock {\em The Semantic Web - ISWC 2015: 14th International Semantic Web
  Conference, Bethlehem, PA, USA, October 11-15, 2015, Proceedings, Part I},
  chapter SPARQL with Property Paths, pages 3--18.
\newblock Springer International Publishing, Cham, 2015.

\bibitem{dataSIOC}
J.~Leskovec, K.~Lang, A.~Dasgupta, and M.~Mahoney.
\newblock Community structure in large networks: Natural cluster sizes and the
  absence of large well-defined clusters.
\newblock {\em Internet Mathematics}, 6(1):29--123, 2009.

\bibitem{LiM03}
Q.~Li and B.~Moon.
\newblock Partition based path join algorithms for xml data.
\newblock In V.~Marík, W.~Retschitzegger, and O.~Stepánková, editors, {\em
  DEXA}, volume 2736 of {\em Lecture Notes in Computer Science}, pages
  160--170. Springer, 2003.

\bibitem{Mabroukeh11}
N.~R. Mabroukeh.
\newblock {\em Semaware: An Ontology-based Web Recommendation System}.
\newblock PhD thesis, University of Windsor, Ontario, Canada, 2011.
\newblock AAINR61939.

\bibitem{Neo4jMan}
T.~N.~T. NeoThechnology.
\newblock The neo4j manual v2.0.0, 2013.

\bibitem{SPARQLSem}
J.~P{\'e}rez, M.~Arenas, and C.~Gutierrez.
\newblock Semantics and complexity of sparql.
\newblock {\em ACM Trans. Database Syst.}, 34(3):16:1--16:45, Sept. 2009.

\bibitem{Robinson}
I.~Robinson, J.~Webber, and E.~Eifrem.
\newblock {\em Graph Databases}.
\newblock O'Reilly Media, Inc., 2013.

\bibitem{Rodriguez15}
M.~A. Rodriguez.
\newblock The gremlin graph traversal machine and language.
\newblock {\em CoRR}, abs/1508.03843, 2015.

\bibitem{SchuhCD16}
S.~Schuh, X.~Chen, and J.~Dittrich.
\newblock An experimental comparison of thirteen relational equi-joins in main
  memory.
\newblock In {\em Proceedings of the 2016 International Conference on
  Management of Data, {SIGMOD} Conference 2016, San Francisco, CA, USA, June 26
  - July 01, 2016}, pages 1961--1976, 2016.

\bibitem{SparqlIncr}
F.~Shmedding.
\newblock Incremental sparql evaluation for query answering on linked data.
\newblock In {\em Second International Workshop on Consuming Linked Data},
  COLD2011, 2011.

\bibitem{SQLGraph}
W.~Sun, A.~Fokoue, K.~Srinivas, A.~Kementsietsidis, G.~Hu, and G.~Xie.
\newblock Sqlgraph: An efficient relational-based property graph store.
\newblock In {\em Proceedings of the 2015 ACM SIGMOD International Conference
  on Management of Data}, SIGMOD '15, pages 1887--1901, New York, NY, USA,
  2015. ACM.

\bibitem{Weichsel}
P.~M. Weichsel.
\newblock The kronecker product of graphs.
\newblock {\em Proceedings of the American Mathematical Society},
  13(1):47–52, 1962.

\bibitem{Yang2015}
J.~Yang and J.~Leskovec.
\newblock Defining and evaluating network communities based on ground-truth.
\newblock {\em Knowledge and Information Systems}, 42(1):181--213, 2015.

\bibitem{Yuan}
P.~Yuan, P.~Liu, B.~Wu, H.~Jin, W.~Zhang, and L.~Liu.
\newblock Triplebit: A fast and compact system for large scale rdf data.
\newblock {\em Proc. VLDB Endow.}, 6(7):517--528, May 2013.

\bibitem{Zeller}
H.~Zeller and J.~Gray.
\newblock An adaptive hash join algorithm for multiuser environments.
\newblock In {\em Proceedings of the 16th International Conference on Very
  Large Data Bases}, VLDB '90, pages 186--197, San Francisco, CA, USA, 1990.
  Morgan Kaufmann Publishers Inc.

\bibitem{Zou09}
L.~Zou, L.~Chen, and M.~T. \"{O}zsu.
\newblock Distance-join: Pattern match query in a large graph database.
\newblock {\em Proc. VLDB Endow.}, 2(1):886--897, Aug. 2009.

\end{thebibliography}

\begin{table*}[!h]
	\caption{Graph Join Running Time with a $\leq$ predicate ($\tau$). GDBMS are tested with different languages: Neo4J is tested with Cypher and RDF4J is tested with SPARQL}\label{tab:evaluateleqjoin}
	\begin{adjustbox}{max width=\textwidth}
		\begin{tabular}{@{}cc|rr|rrr|rr@{}}
			\toprule
			\multicolumn{2}{c}{\textbf{Operands Size ($|V|$)}} &
			\multicolumn{2}{c}{\textbf{Result}} & \multicolumn{3}{c}{\textbf{Join Time} $\tau$(ms)}  & \multicolumn{2}{c}{\textbf{Speed-up}} \\
			Left & Right & Size ($|V|$) & Avg. Multiplicity & Cypher-Neo4J ($\tau_N$) & RDF4J-SPARQL ($\tau_R$) & Our CoGrouped Join ($\tau'$) & Cypher ($\sfrac{\tau_N}{\tau'}$) & SPARQL ($\sfrac{\tau_R}{\tau'}$) \\
			\midrule
			\begin{filecontents*}{calcleq.csv}
"size","sizer","avg","neo4j","rdf4j","proposed","spedA","spedB"
$10$,55,5.5,194.94,10.49,0.40,487.35,26.22
$10^2$,5130,51.3,1\,082.58,556.95,3.83,282.66,145.42
$10^3$,510\,055,509.54,288\,619.91,88\,312.93,558.73,516.56,158.06
\end{filecontents*}
\csvreader[late after line=\\]{calcleq.csv}{}{\csvcoli & \csvcoli & \csvcolii & \csvcoliii & \csvcoliv & \csvcolv  & \csvcolvi& \csvcolvii & \csvcolviii}%
			\bottomrule
		\end{tabular}
	\end{adjustbox}
\end{table*}
\begin{algorithm}[!th]
	\caption{CoGrouped Graph Conjunctive LessEqual-Join}\label{alg:cogroupedleq}
	\begin{algorithmic}[1]
		\Procedure{LeqJoin}{$G_1,G_2,A_1,B_j$}\Comment{$G_1\Join_{A_i\leq B_j}^{\wedge} G_2$}
		\State $V\leftarrow \emptyset;\; E\leftarrow \emptyset$
		\State $HLeft \leftarrow \texttt{iterate\_from\_min}(V_1)$
		\State $HRight \leftarrow \texttt{iterate\_from\_max}(V_2)$
		\For{\textbf{each} $h_l\in HLeft$}
		\For{\textbf{each} $u\in V_1$ \textbf{s.t.} $h(u)=h_l$}
		\For{\textbf{each} $h_r\in HRight$}
		\If{$h_l>h_r$} \textbf{break}
		\EndIf 
		\For{\textbf{each} $v\in V_2$ \textbf{s.t.} $h(v)=h_r$}
		\If{$u.A_i\leq v.B_j, (u\oplus v)[A]=u,$\par 
			\hskip\algorithmicindent\hskip\algorithmicindent\hskip\algorithmicindent\hskip\algorithmicindent\hskip\algorithmicindent $(u\oplus v)[B]=v$}
		\For{\textbf{each} $nu\in out_{V_1}(u)$}
		\For{\textbf{each} $nv\in out_{V_2}(v)$}
		\If{$\theta(nu,nv),$\par 
			\hskip\algorithmicindent\hskip\algorithmicindent\hskip\algorithmicindent\hskip\algorithmicindent\hskip\algorithmicindent\hskip\algorithmicindent\hskip\algorithmicindent\hskip\algorithmicindent $(nu\oplus nv)[A]=nu,$\par 
			\hskip\algorithmicindent\hskip\algorithmicindent\hskip\algorithmicindent\hskip\algorithmicindent\hskip\algorithmicindent\hskip\algorithmicindent\hskip\algorithmicindent\hskip\algorithmicindent$(nu\oplus nv)[B]=nv$}
		\State {$V\leftarrow V\cup \{nu\oplus nv\}$}
		\State {$E\leftarrow E\cup \{(u\oplus v, nu\oplus nv)\}$}
		\EndIf
		\EndFor
		\EndFor
		\EndIf
		\EndFor
		\EndFor
		\EndFor
		\EndFor
		\EndProcedure
	\end{algorithmic}
\end{algorithm}

\begin{algorithm*}[!h]
	\caption{Hash Join algorithm using the vectorial data model in Secondary Memory and sequential reads}\label{alg:hashBulk}
	\label{alg:proposed}
	\centering
	\scalebox{0.826}{
		\begin{minipage}{1.2\linewidth}
	\begin{algorithmic}[1]
		\Function{Offset}{Index,VA,i}
		\If{i$<|$Index$|-1$}
		\State \Return{VA[Index[i].VAOffset,$\dots$,Index[i+1].VAOffset]}
		\Else
		\State \Return{VA[Index[i].VAOffset,$\dots$,$|$\textit{VA}$|$]}
		\EndIf
		\EndFunction
		
		\Function{readVertex}{Array};
		\Comment{Returns the first vertex stored inside Array}
		\EndFunction
		\Function{V$_X$[$i$]}{}:=
		\textsc{readVertex}(\textsc{Offset}(VertexIndex$_X$,VA$_X$,$i$)) \Comment{$X$ is a placeholder for either left or right}
		\EndFunction
		\State
		
		\Function{GraphHashJoinVisit}{$G_L,G_R,\theta$} \Comment{$G_X\equiv($VertexIndex$_X$,VA$_X$,Hash$_X,A_X)$}
		\State bulkGraph:=\textbf{new} Stack()  \Comment{Creates the Bulk Graph}
		\State $\theta'(u,v):=\theta(u,v)\wedge (u\oplus v)[A_1]=u \wedge (u\oplus v)[A_2]=v$
		\Comment{Extends the $\theta$ equivalence predicate with the join test}
		\State{HI:=Hash$_L\cap$Hash$_R$}\Comment{Get the hash values in common to the two graphs}
		\For{\textbf{each} $h_c\in $HI}
		\State Array$_L$ := \textsc{Offset}(Hash$_L$,VA$_L$,$h^L_c$) \Comment{$h^X_c$ returns the index of hash $h_c$ in Hash$_X$}
		
		\While{$|$Array$_L|>0$} \Comment{Checks if there are more vertices to read in Array$_L$}
		\State{u:=\textsc{readVertex}(Array$_L$)}
		\State{Array$_R$:= \textsc{Offset}(Hash$_R$,VA$_R$,$h^R_c$)}
		\While{$|$Array$_R|>0$} \Comment{Checks if there are more vertices to read in Array$_R$}
		\State{v:=\textsc{readVertex}(Array$_R$)}
		\If{(!$\theta'$(u,v))} \textbf{continue}
		\EndIf
		\State{np:=u$\oplus $v}
		\For{\textbf{each} nup$\in$u.out} \Comment{u.out is the array containing the indices of u's outgoing neighbours}
		\State{nu:=V$_L$[nup]} \Comment{Retrieves the actual vertex}
		\If{$h($nu$)\notin $HI} \textbf{continue}
		\EndIf
		\For{\textbf{each} nvp$\in$v.out} \Comment{v.out is the array containing the indices of v's outgoing neighbours}
		\State{nv:=V$_R$[nvp]} \Comment{Retrieves the actual vertex}
		\If{$h($nv$)\notin $HI} \textbf{continue}
		\EndIf
		\If{(!$\theta'($nu,nv$)$)} \textbf{continue}
		\EndIf
		\State{np.neighborus.add(\,(nup,nvp)\,)}
		\EndFor
		\EndFor
		\State{bulkGraph.push(np)}
		\State{Array$_R$:=Array$_R$[\,$|$v$|$,$\dots$,$|$Array$_R|$\,]} \Comment{Prepare the array to read eventually the next vertex}
		\EndWhile 
		\State{Array$_L$:=Array$_L$[\,$|$u$|$,$\dots$,$|$Array$_L|$\,]} \Comment{Prepare the array to read eventually the next vertex}
	
		\EndWhile 
		\EndFor
		\Return{bulkGraph}
		\EndFunction
	\end{algorithmic}
\end{minipage}
}
\end{algorithm*}
\appendix
\section{CoGrouped Graph Conjunctive LessEqual-Join}\label{app:leq}
Algorithm \ref{alg:cogroupedleq} shows the implementation of the graph join when the predicate is a 
less-equal over one attribute. 
Given a less-equal predicate $\theta(u,v)=u.A_i\leq v.B_j$ and an hashing function $h$ such that
$h(u)\leq h(v)$ then $u.A_i\leq v.B_j$, 
we join all the vertices from the left graph that have a vertex hash code that is less or equal 
of the left one. 
Since the hashes can be scanned in advance through a linear data structure, we choose to 
scan the left operands' hashes in ascending order: for each new hash value of such operand, we
scan the right operands' hashes in descending order. In this way, if the left operand's current hash
is greater than the one currently scanned over the right graph, we can break the whole
internal iteration and proceed with the next hash for the left operand.
Then, the algorithm behaves similarly when we have to join the neighbour vertices.

In order to compare our results with the EquiJoin ones, we performed our
benchmarking over the same experimental conditions as described in Section \ref{ref:results}.

The hash scanning choice is revealed to be poignant by the benchmarks that have been carried out. 
Table \ref{tab:evaluateleqjoin} provides the results for the less-equal test, where we kept the same 
subgraph samples that were used for the equi-join algorithm. In this case all the benchmarked implementations
stopped with a graph input size of $10^4$ due to the high multiplicity of the result (with a $10^3$ dataset, we 
reach an average multiplicity of $\approx 509$). In this case, SPARQL evaluated with RDF4J has a better
performance than Cypher evaluated over Neo4J: our algorithm is at most two orders of magnitude faster
than the evaluation of both Cypher query with Neo4J and of the SPARQL query over RDF4J. 

As a consequence, we confirm that our algorithm has better performances than current implementations
when the multiplicity coefficient increases. We
propose to store the graph results in secondary memory for further implementations. In this way we allow the algorithm
to finish the join computation with no thwarts caused by the primary-memory size limitations.

\section{Implementing the CoGrouping Graph Conjunctive EquiJoin}\label{spec:implement}

Algorithm \ref{alg:proposed} assumes a hash-ordered vertex 
representation through which a CoGroup-join algorithm over the vertices is possible. 
This CoGroup-join technique (lines 15-22) is based on a previous evaluation of the hash value
of each vertex, and hence this reduces the search cost of finding vertices with 
similar features(lines 13-14). Such technique is extended with the visit of the 
neighbour vertices (lines 23-29), where the neighbours' hash value is used as 
a heuristic (line 25 and line 28) in order to skip a test of the equi-join predicate $\theta$
returning false. The result is stored using a Bulk Graph. 
\balance

\end{document}